\documentclass[10pt,twocolumn]{article}

\usepackage[a4paper,top=2cm,bottom=2.2cm,left=1.6cm,right=1.6cm,columnsep=0.6cm]{geometry}
\usepackage[T1]{fontenc}
\usepackage{lmodern}
\usepackage{microtype}
\usepackage{graphicx}
\usepackage{natbib}
\bibpunct{(}{)}{;}{a}{}{,}
\usepackage{authblk}

\usepackage[font=small,labelfont=bf,labelsep=period]{caption}
\usepackage{titlesec}

\titleformat*{\section}{\large\bfseries}
\titleformat*{\subsection}{\normalsize\bfseries}
\usepackage{amsmath,amssymb}
\usepackage{booktabs}
\usepackage{xcolor}
\usepackage{tikz}
\usetikzlibrary{patterns,decorations.pathreplacing}
\usepackage{siunitx}
\usepackage[colorlinks,allcolors=blue]{hyperref}

\makeatletter
\newcommand{\keepspace}[1]{\par\penalty-100\begingroup\dimen@=#1\relax
  \dimen@ii\pagegoal\advance\dimen@ii-\pagetotal
  \ifdim\dimen@>\dimen@ii\ifdim\dimen@ii>\z@\vfil\fi\break\fi\endgroup}
\let\gs@section\section
\renewcommand{\section}{\keepspace{6\baselineskip}\gs@section}
\makeatother

\newcommand{\pkg}[1]{{\normalfont\scshape #1}}
\newcommand{\gradsolve}{\pkg{gradsolve}}
\newcommand{\diffrax}{\pkg{diffrax}}
\newcommand{\jax}{\pkg{jax}}
\newcommand{\scipy}{\pkg{scipy}}
\newcommand{\gala}{\pkg{gala}}
\newcommand{\precessionpkg}{\pkg{precession}}
\newcommand{\code}[1]{\texttt{#1}}

\newcommand{\stRefTol}{\num{1e-13}}
\newcommand{\prRefTol}{\num{1e-13}}
\newcommand{\deRefRtol}{\num{1e-11}}
\newcommand{\deRefAtol}{\num{1e-13}}

\newcommand{\cpuModel}{AMD EPYC 9745}
\newcommand{\cpuNodeSpeedupHi}{95}
\newcommand{\cpuNodeSpeedupLo}{93}
\newcommand{\deAtolRatio}{\num{1e-3}}
\newcommand{\deBudget}{\num{1e-4}}
\newcommand{\deDesiMinErr}{\num{2.5e-3}}
\newcommand{\deErrPnn}{\num{7.9e-5}}
\newcommand{\deFrozenEzero}{0.82}
\newcommand{\deFrozenOmShift}{11}
\newcommand{\deFwdDfxArm}{Dopri5, \code{rtol} = \num{1.78e-4}}
\newcommand{\deFwdDfxArmFull}{Tsit5, \code{rtol} = \num{3.16e-4}}
\newcommand{\deFwdDfxCondEfive}{0.077}
\newcommand{\deFwdDfxCondEfour}{0.35}
\newcommand{\deFwdDfxCondEsix}{0.066}
\newcommand{\deFwdDfxFullEfive}{0.52}
\newcommand{\deFwdDfxFullEfour}{2.1}
\newcommand{\deFwdDfxFullEsix}{0.56}
\newcommand{\deFwdFactorCondEfive}{5.9}
\newcommand{\deFwdFactorCondEfour}{3.1}
\newcommand{\deFwdFactorCondEsix}{18}
\newcommand{\deFwdFactorFullEfive}{4.7}
\newcommand{\deFwdFactorFullEfour}{2.6}
\newcommand{\deFwdFactorFullEsix}{11}
\newcommand{\deFwdGsCondEfive}{0.013}
\newcommand{\deFwdGsCondEfour}{0.11}
\newcommand{\deFwdGsCondEsix}{0.0037}
\newcommand{\deFwdGsFullEfive}{0.11}
\newcommand{\deFwdGsFullEfour}{0.81}
\newcommand{\deFwdGsFullEsix}{0.051}
\newcommand{\deFwdScipyCond}{520}
\newcommand{\deNodeSpeedupGsEsix}{\num{1500}}
\newcommand{\deNover}{\num{1573}}
\newcommand{\deNz}{\num{7}}
\newcommand{\deNzDense}{\num{64}}
\newcommand{\deObsArmsMax}{\num{2.6e-2}}
\newcommand{\deObsDchiZero}{4.6}
\newcommand{\deObsIndepMax}{\num{4.1e-2}}
\newcommand{\deObsLamBest}{0.80}
\newcommand{\dePriorLam}{[0, 1.5]}
\newcommand{\dePriorOm}{[0.28, 0.34]}
\newcommand{\deProfCheckMax}{\num{4.1e-2}}
\newcommand{\deProfConv}{$|\Delta\chi^2| < \num{1e-4}$ over the last step and a Gauss--Newton decrement below \num{1e-6}}
\newcommand{\deProfDfxCond}{2.5}
\newcommand{\deProfDfxFull}{4.1}
\newcommand{\deProfDfxSolveCond}{1.8}
\newcommand{\deProfDfxSolveFull}{3.5}
\newcommand{\deProfDthetaMax}{\num{1.3e-3}}
\newcommand{\deProfFactorCond}{3.0}
\newcommand{\deProfFactorFull}{3.5}
\newcommand{\deProfGrid}{\num{100}}
\newcommand{\deProfGsCond}{0.83}
\newcommand{\deProfGsFull}{1.2}
\newcommand{\deProfGsSolveCond}{0.17}
\newcommand{\deProfGsSolveFull}{0.52}
\newcommand{\deProfIters}{\num{20}}
\newcommand{\deProfMocks}{\num{100}}
\newcommand{\deProfN}{\num{10000}}
\newcommand{\deProfNcheck}{\num{5}}
\newcommand{\deProfNodes}{\num{64}}
\newcommand{\deProfPostCond}{0.66}
\newcommand{\deProfPostFull}{0.66}
\newcommand{\deProfScipyCond}{3000}
\newcommand{\deProfScipyFull}{11000}
\newcommand{\deProfSolveFactorCond}{11}
\newcommand{\deProfSolveFactorFull}{6.7}
\newcommand{\deRefDiffMax}{\num{1e-6}}
\newcommand{\deRefN}{\num{10000}}
\newcommand{\deZstart}{\num{50}}
\newcommand{\hwCPU}{an AMD EPYC 9745 CPU (128 cores, up to 3.7\,GHz, 256\,MB L3 cache)}
\newcommand{\prBudget}{\num{2e-7}}
\newcommand{\prCpuMs}{220}
\newcommand{\prDfxArm}{Verner~7, \code{rtol} = \num{3.16e-9}}
\newcommand{\prDfxDrawn}{34}
\newcommand{\prFactorTsitDrawn}{4.2}
\newcommand{\prFactorVernDrawn}{11}
\newcommand{\prN}{\num{1000000}}
\newcommand{\prNodeSpeedupGsEsix}{\num{760}}
\newcommand{\prParityN}{\num{20000}}
\newcommand{\prRefDiffMax}{\num{9.1e-4}}
\newcommand{\prRelErrPnn}{\num{1.3e-7}}
\newcommand{\prStepsDfx}{\num{901}}
\newcommand{\prStepsGs}{\num{911}}
\newcommand{\prTsitDrawn}{8.1}
\newcommand{\prTsitErrPnn}{\num{1.6e-7}}
\newcommand{\prTsitNover}{\num{1791}}
\newcommand{\prTsitOldErrPnn}{\num{2.8e-7}}
\newcommand{\prTsitOldRtol}{\num{1e-9}}
\newcommand{\prTsitRtol}{\num{5.62e-10}}
\newcommand{\prVernDrawn}{3.2}
\newcommand{\prVernErrPnn}{\num{1.8e-7}}
\newcommand{\prVernNover}{\num{4827}}
\newcommand{\prVernRtol}{\num{3.16e-9}}
\newcommand{\stCpuUs}{170}
\newcommand{\stFwdBudget}{\num{1e-4}}
\newcommand{\stFwdDfxArm}{Tsit5, \code{rtol} = \num{3.16e-7}}
\newcommand{\stFwdDfxArmEfour}{Dopri8, \code{rtol} = \num{1e-6}}
\newcommand{\stFwdDfxDrawnEfive}{2.0}
\newcommand{\stFwdDfxDrawnEfour}{6.2}
\newcommand{\stFwdDfxDrawnEsix}{1.2}
\newcommand{\stFwdDfxNover}{\num{236}}
\newcommand{\stFwdErrPnn}{\num{5.7e-5}}
\newcommand{\stFwdFactorDrawnEfive}{15}
\newcommand{\stFwdFactorDrawnEfour}{13}
\newcommand{\stFwdFactorDrawnEsix}{11}
\newcommand{\stFwdGsDrawnEfive}{0.13}
\newcommand{\stFwdGsDrawnEfour}{0.49}
\newcommand{\stFwdGsDrawnEsix}{0.12}
\newcommand{\stFwdNover}{\num{573}}
\newcommand{\stGradBudget}{\num{1e-3}}
\newcommand{\stGradDfxArm}{Tsit5 on the sensitivity equations, \code{rtol} = \num{3.16e-7}}
\newcommand{\stGradDfxArmEfour}{Dopri8 through \code{jax.jacfwd}, \code{rtol} = \num{3.16e-6}}
\newcommand{\stGradDfxDrawnEfive}{4.4}
\newcommand{\stGradDfxDrawnEfour}{8.1}
\newcommand{\stGradDfxDrawnEsix}{3.1}
\newcommand{\stGradDfxNover}{\num{33}}
\newcommand{\stGradErrPnn}{\num{6e-4}}
\newcommand{\stGradFactorDrawnEfive}{19}
\newcommand{\stGradFactorDrawnEfour}{9.2}
\newcommand{\stGradFactorDrawnEsix}{15}
\newcommand{\stGradGsDrawnEfive}{0.24}
\newcommand{\stGradGsDrawnEfour}{0.88}
\newcommand{\stGradGsDrawnEsix}{0.22}
\newcommand{\stGradNover}{\num{21}}
\newcommand{\stGradSameDrawnEfive}{4.4}
\newcommand{\stGradSameDrawnEfour}{14}
\newcommand{\stGradSameDrawnEsix}{3.1}
\newcommand{\stGradSameFactorDrawnEfive}{19}
\newcommand{\stGradSameFactorDrawnEfour}{16}
\newcommand{\stGradSameFactorDrawnEsix}{15}
\newcommand{\stNPeriods}{a median of 3.7}
\newcommand{\stNodeSpeedupGsEsix}{\num{8.5}}
\newcommand{\stRefDiffMax}{\num{1.2e-7}}
\newcommand{\stRtol}{\num{3.16e-7}}
\newcommand{\stStepsDfx}{\num{176}}
\newcommand{\stStepsGs}{\num{171}}

\begin{document}

\title{\bfseries Differentiable astrophysics at scale:\\
solving and differentiating ODE ensembles on the GPU}

\author{Alessio Spurio Mancini\thanks{E-mail: \href{mailto:alessio.spuriomancini@rhul.ac.uk}{alessio.spuriomancini@rhul.ac.uk}}}
\affil{Department of Physical Sciences and Engineering, Royal Holloway, University of London,\\
Egham Hill, Egham, TW20 0EX, UK}
\date{}

\twocolumn[
\begin{@twocolumnfalse}
\maketitle
\begin{abstract}
Astronomers increasingly fit their models with gradient-based methods, such as Hamiltonian Monte
Carlo, which need the derivatives of the model with respect to its parameters. In many analyses a prediction requires solving a small system of
ordinary differential equations (ODEs) for thousands to millions of parameter sets, and this
ensemble of integrations often sets the cost of the analysis. We introduce to the astronomical
community \gradsolve{}, a \jax{} library that moves this computation to graphics processing units
(GPUs): it integrates each member of the ensemble in its own GPU thread with its own step size and
returns the derivatives with respect to the parameters in the same pass. We measure the speed-up it provides in three
examples from different fields of astronomy, stellar orbits in the Galactic potential, the
expansion history of a dark-energy cosmology and the spin precession of binary black holes, with
every code held to the same accuracy requirement. For a million trajectories \gradsolve{} runs
\stNodeSpeedupGsEsix{} to \deNodeSpeedupGsEsix{} times faster than the serial CPU code of each example
running on all 128 cores of a CPU, and several orders of magnitude faster than on one core. On the same GPU it is also
\deFwdFactorFullEsix{} to \stGradFactorDrawnEsix{} times faster than \diffrax{}, the
state-of-the-art ODE library in \jax{}, in all three examples. With \gradsolve{} a million such integrations take seconds or less on one GPU, so
gradient-based analyses that need ensembles of this size become routine. The code is publicly available at
\url{https://github.com/ECLIPSE-AI4Science/gradsolve}.

\end{abstract}
\bigskip
\end{@twocolumnfalse}
]

\section{Introduction}
\label{sec:intro}
\begin{figure*}[t]
\centering
\includegraphics[width=\textwidth]{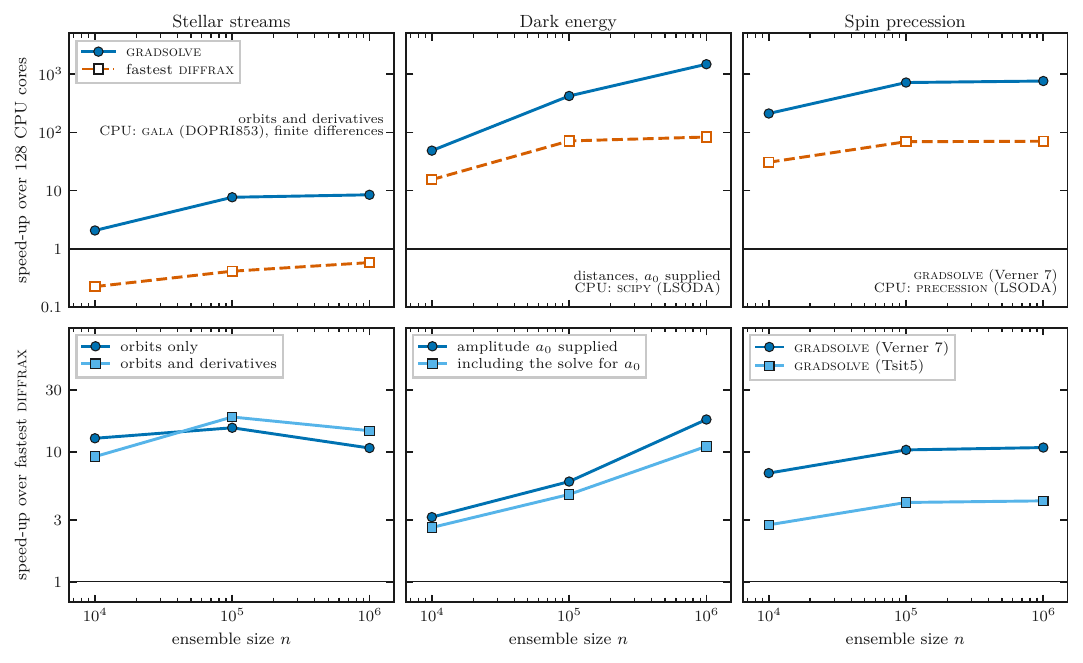}
\caption{Speed-up of \gradsolve{} against ensemble size $n$, for stellar-stream orbits with their
derivatives with respect to $(v_c, q)$ (left; Section~\ref{sec:streams}), dark-energy distances
(middle; Section~\ref{sec:darkenergy}) and spin precession (right; Section~\ref{sec:precession}). \emph{Top}: speed-up over the serial CPU code of each example running on all 128 cores of an
\cpuModel{} CPU (one process per core, measured), the CPU cost per trajectory divided by the GPU cost
per trajectory, for \gradsolve{} (solid) and for the fastest accepted \diffrax{} configuration \citep{Kidger2021} (dashed), both on the same GPU; the horizontal line marks equal cost, so a curve above it beats the
whole CPU. The precession panel shows \gradsolve{} (Verner~7). Against a single core the same ratios are larger by a further \cpuNodeSpeedupLo{} to
\cpuNodeSpeedupHi{} times. The CPU codes are \gala{} (DOPRI853; \citealt{PriceWhelan2017gala}), compiled C with derivatives by
finite differences, and \scipy{} (LSODA; \citealt{Virtanen2020scipy, Hindmarsh1983odepack,
Petzold1983lsoda}) and \precessionpkg{} (LSODA; \citealt{GerosaKesden2016precession, Gerosa2023precessionv2}),
both with a right-hand side written in Python.
These ratios combine differences in hardware and software. \emph{Bottom}: speed-up of \gradsolve{}
over the fastest accepted \diffrax{} configuration on the same GPU and ensemble, for two quantities
per example: orbits alone and with derivatives; distances with the amplitude $a_0$ supplied and
including the solve for it; and \gradsolve{} (Verner~7) and \gradsolve{} (Tsit5) for precession. The
horizontal line marks equal cost. Every code, CPU codes included, meets the accuracy requirement of its example
(Section~\ref{sec:comparison}); the full timings are in Tables~\ref{tab:streams}, \ref{tab:precession}
and~\ref{tab:de_forward} in the appendix.}
\label{fig:speedup}
\end{figure*}

Gradient-based methods are now common in astrophysical inference. Optimisers \citep{Liu1989lbfgs, Kingma2015adam}, Fisher
forecasts \citep[e.g.][]{Tegmark1997fisher, JaxCosmo2023, Iacovelli2022gwfast} and Hamiltonian Monte Carlo
\citep{Duane1987hmc, Neal2011hmc, Hoffman2014nuts}, available in \jax{} libraries
\citep[e.g.][]{Cabezas2024blackjax, Phan2019numpyro}, need the derivatives of a model prediction with
respect to its parameters, and automatic differentiation \citep[see][]{GriewankWalther2008,
Baydin2018ad}, most often in \jax{} \citep{Bradbury2018jax}, supplies them. Together with emulators and graphics
processing units (GPUs), these gradients have made Bayesian inference over many parameters practical
\citep[e.g.][]{JaxCosmo2023, CosmoPowerJAX2023, RuizZapatero2024limberjack, Piras2024future}. Differentiable models now
exist in many areas of astrophysics, and they differ in where their cost lies. At one end a single
large simulation is differentiated as a whole, as in field-level inference of the cosmic density
field \citep[e.g.][]{Jasche2013borg, Wang2014elucid, Jasche2015sdss, Seljak2017, Porqueres2021shear} with differentiable particle-mesh
simulations \citep[e.g.][]{Modi2021flowpm, Li2024pmwd}, in differentiable $N$-body models of stellar
streams \citep{Viterbo2025odisseo} and in hydrodynamics
\citep[e.g.][]{Horowitz2025diffhydro, Storcks2024jf1uids}; the parallel work
lies within one simulation, and derivatives with respect to an initial field with one value per grid
cell call for reverse-mode (adjoint) differentiation \citep[e.g.][]{Li2024pmwd}. At the other end the model is
an explicit function of its parameters or an emulator of an expensive code. Examples are emulators of
cosmological power spectra \citep[e.g.][]{Heitmann2009coyote, Arico2021bacco, Mootoovaloo2022kernel,
SpurioMancini2022cosmopower, Nygaard2023connect, Bonici2024capse} and of intermediate quantities such as
CMB source functions \citep[e.g.][]{Albers2019cosmicnet}, auto-differentiable cosmology
libraries, likelihoods and forward models \citep[e.g.][]{JaxCosmo2023, RuizZapatero2024limberjack,
Balkenhol2024candl, Kern2025, Reymond2026swiftcl, HIcosmo2026}, gravitational waveforms and the inference built on them
\citep[e.g.][]{Edwards2024ripple, Wong2023flowMC, Wong2023jim}, exoplanet light curves and spectra
\citep[e.g.][]{ForemanMackey2021exoplanet, Luger2019starry, Kawahara2022exojax}, stellar and galaxy
spectra \citep[e.g.][]{Wheeler2023korg, Wheeler2024korg, Alsing2020speculator, Hearin2023dsps}, halo growth
\citep{Hearin2021diffmah} and strong lenses \citep[e.g.][]{Chianese2020lensing,
Gu2022gigalens, Galan2022herculens, Stone2024caustics}. Between the two lie models in which every
evaluation solves a small system of ordinary differential equations (ODEs), repeated for every
cosmology a sampler visits, every star of a stream model or every binary of a synthetic population.
Examples are nucleosynthesis \citep{Giovanetti2024linx}, neutron-star structure
\citep{Kacprzak2025jester} and stellar-stream models \citep{Alvey2023albatross,
Nibauer2024}; differentiable Einstein--Boltzmann solvers \citep{DiscoDJ2023,
Sletmoen2026symboltz, Nguyen2026clax, Zhou2026abcmb} solve larger, stiff systems of this kind.

An adaptive solver chooses each step length from an estimate of the error made in that step, so
the trajectories of one equation take different numbers of steps. \diffrax{}\footnote{\url{https://github.com/patrick-kidger/diffrax}} \citep{Kidger2021},
the state-of-the-art library for solving ODEs in \jax{}, advances a vectorised ensemble as one array,
attempting a step for every trajectory until the last one finishes, so its cost follows the largest
step count rather than the mean (Figure~\ref{fig:mechanism}). GPU ensemble solvers written in CUDA and Julia integrate each trajectory in its own GPU thread with its own step size
\citep[e.g.][]{Niemeyer2014gpu, Nagy2022mpgos, Rackauckas2023diffeqgpu}, and \pkg{torchode} keeps one step
size per problem in \pkg{pytorch} \citep{Lienen2022torchode}; these run outside the automatic
differentiation of \jax{}. A different route replaces the numerical integrator by a neural network trained
to satisfy the equations \citep[physics-informed neural networks; e.g.][]{Raissi2019pinn,
Protopapas2026pinnbook}; here we keep the adaptive integrator, whose error is controlled step by step.
\gradsolve{} \citep{SpurioMancini2026gradsolve} generates such a per-thread kernel from a
right-hand side written in \jax{}. The version used here also interpolates the solution at the
requested times and integrates the forward sensitivities, the derivatives of the state with respect
to the parameters, inside the same kernel (Section~\ref{sec:sensitivities}); in the library paper
both come from a separate differentiable replay of the recorded steps in \jax{}. Forward
sensitivities cost less than reverse-mode (adjoint) differentiation \citep{Cao2003adjoint,
Chen2018neuralode} when few parameters are differentiated and many outputs need derivatives
\citep[see][]{Hindmarsh2005sundials, Ma2021sensitivity}.

We introduce \gradsolve{} to the astronomical community to accelerate such analyses, and apply it
to three problems: stellar orbits in a flattened Galactic halo, differentiated at one fixed
potential with respect to two parameters shared by all stars, as in stream models of the Milky Way
(Section~\ref{sec:streams}); the expansion history of a flat universe with thawing quintessence,
for many cosmologies and inside a profile likelihood (Section~\ref{sec:darkenergy}); and spin
precession across a population of binary black holes (Section~\ref{sec:precession}).
Figure~\ref{fig:speedup} summarises the speed-up the library provides in each of them, in two steps. Moving the
integration from a serial CPU code to the GPU makes it \stNodeSpeedupGsEsix{} to
\deNodeSpeedupGsEsix{} times faster than that code running on all 128 cores of the machine, for a
million trajectories, and several orders of magnitude faster than one core (top row); this gain
combines hardware and software. On the same GPU, \gradsolve{} is in addition about an order of magnitude faster than
\diffrax{} for large ensembles, at the same accuracy (bottom row).
Section~\ref{sec:method} describes the method and how the two codes are compared,
Sections~\ref{sec:streams}--\ref{sec:precession} the applications, and
Sections~\ref{sec:discussion} and~\ref{sec:conclusions} discuss where the approach applies and
conclude.

\section{Method}
\label{sec:method}

This section describes the computation that the three applications share, how \gradsolve{} carries
it out, and how we compare it with \diffrax{}. The library and its reverse-mode gradient are
described by \citet{SpurioMancini2026gradsolve}; here we also use two quantities that the same kernel computes, the solution interpolated at requested times and the forward sensitivities. The numerical methods are
standard \citep[see e.g.][]{HairerNorsettWanner1993, HairerWanner1996}, and
Appendix~\ref{sec:app-numerics} gives the equations behind this section.

\subsection{The ensemble problem}
\label{sec:problem}

An ensemble is a set of $n$ independent initial-value problems with the same right-hand side $f$,
\begin{equation}
  \frac{\mathrm{d}y_i}{\mathrm{d}t} = f(t, y_i, \theta_i), \qquad
  y_i(t_0) = y_{0,i}, \qquad i = 1, \dots, n,
  \label{eq:ivp}
\end{equation}
with state $y_i(t) \in \mathbb{R}^d$ and parameters $\theta_i \in \mathbb{R}^p$. The state is the
position and velocity of a star in Section~\ref{sec:streams} ($d=6$); a scalar field, its
derivative and a distance integral of one cosmology in Section~\ref{sec:darkenergy} ($d=3$, with the
number of e-folds as $t$); and the directions of the orbital angular momentum and the two spins of
one binary in Section~\ref{sec:precession} ($d=9$, with the orbital velocity as $t$).

An analysis reads the solution at output times $t_k$ set by the data rather than by the solver. A
gradient-based analysis also needs the sensitivities $S_i(t_k) = \partial y_i(t_k)/\partial\theta$,
where $\theta$ now denotes the $m \le p$ parameters that the analysis varies, two in each of the two
examples that need derivatives. The gradient of a likelihood or a $\chi^2$ built from the outputs
then follows from these sensitivities by the chain rule (equation~\ref{eq:chain}), which automatic
differentiation applies to the code that builds it. Almost all of the cost therefore lies in
computing the solutions and their sensitivities.

\subsection{Adaptive integration}
\label{sec:adaptive}
\label{sec:primer}

The three applications use explicit Runge--Kutta methods. Each step of length $h$ evaluates $f$ a
fixed number of times and combines the evaluations in two ways: one gives the new state and the
other, at no extra cost, an estimate of the error made in that step. The user sets a relative
tolerance \code{rtol} and an absolute tolerance \code{atol}. A step is accepted when its error estimate, divided component by component by $\code{atol} + \code{rtol}\,|y|$ and averaged over the components of the state, is smaller than one, and is otherwise retried with a shorter step; after every attempt
the next step length is chosen from the size of the estimate (equations~\ref{eq:errnorm}
and~\ref{eq:controller}). The step length therefore follows the solution, and trajectories of the
same equation take different numbers of steps. A rejected attempt costs as much as an accepted one,
so we count all step attempts, $N_i$ for trajectory $i$.

We use Tsit5 \citep{Tsitouras2011} and Dopri5 \citep{DormandPrince1980}, of fifth order, Verner~7
\citep{Verner2010} and Dopri8 \citep{PrinceDormand1981}. Higher-order methods take fewer but more
expensive steps and are more efficient at tight tolerances. The tolerances bound the estimated error
of each step, not the error of the solution, so every method is judged by the error it reaches
against a converged reference (Section~\ref{sec:comparison}).

These explicit methods suit \emph{non-stiff} problems. A problem is \emph{stiff} when some perturbation of its
solution decays much faster than the solution itself changes, as in chemical networks or the tightly
coupled photon--baryon fluid of the Einstein--Boltzmann equations \citep[e.g.][]{MaBertschinger1995,
Lewis2000camb, Blas2011class}. An explicit method is then stable only for steps as short as that
decay, so stability rather than accuracy sets the step length; implicit and Rosenbrock methods avoid
this, and \gradsolve{} provides a Rosenbrock kernel, benchmarked by \citet{SpurioMancini2026gradsolve}
on two standard stiff test problems \citep{Robertson1966, Schafer1975hires}. In each of the three problems here the shortest time-scale (the
orbital period, the expansion time, the precession period) is one that the solution itself follows,
so steps chosen for accuracy are also stable and the explicit methods apply.

The output times $t_k$ generally fall inside steps, and ending a step at each of them would shorten
the steps. Instead, the evaluations of an accepted step define a polynomial that gives the solution
anywhere inside the step (equation~\ref{eq:dense}). With this dense output the steps do not depend
on the output times. For Tsit5 each output costs one evaluation of the polynomial.
The dark-energy distances are read from the Tsit5 polynomial.

\subsection{Derivatives}
\label{sec:sensitivities}

Differentiating equation~(\ref{eq:ivp}) with respect to $\theta$ gives the forward sensitivity
equations,
\begin{equation}
  \frac{\mathrm{d}S_i}{\mathrm{d}t} = \frac{\partial f}{\partial y}\, S_i
  + \frac{\partial f}{\partial\theta}, \qquad
  S_i(t_0) = \frac{\partial y_{0,i}}{\partial\theta},
  \label{eq:sens}
\end{equation}
with the derivatives of $f$ evaluated along the trajectory. \gradsolve{} integrates these equations
together with $y_i$ as one system of $d(1+m)$ components, and forward-mode automatic differentiation of $f$ supplies their right-hand
side, so the user writes only $f$. The step length is set by the error of $y_i$ alone, so the steps
are those of the solve without derivatives, and the dense output gives $S_i(t_k)$ as well as
$y_i(t_k)$. On a given sequence of steps the result is exactly the derivative of the numerical
solution (equation~\ref{eq:stage-sens}), which is also what forward-mode differentiation through
\diffrax{} returns, since \diffrax{} holds its step lengths fixed when it differentiates
\citep[\S5.4.2.2]{Kidger2021}.

One integration of equation~(\ref{eq:sens}) gives the derivatives of every output of a trajectory,
at roughly $1+m$ times the cost of the solve alone. Reverse-mode, or adjoint, differentiation
\citep[see e.g.][]{Cao2003adjoint, Chen2018neuralode} scales the other way: one backward
integration gives the gradient of one scalar with respect to any number of parameters, but it needs
the forward solution stored or recomputed (equations~\ref{eq:adjoint} and~\ref{eq:disc-modes}).
Forward sensitivities therefore suit analyses with few parameters and many outputs, such as the
least-squares fits and Fisher forecasts of this paper, and since they store nothing between steps,
each of millions of trajectories needs only the memory of its own GPU thread. For a sampler that
needs the gradient of the log-likelihood with respect to many parameters, \gradsolve{} provides a
reverse-mode gradient \citep{SpurioMancini2026gradsolve}.

\subsection{One trajectory per GPU thread}
\label{sec:library}

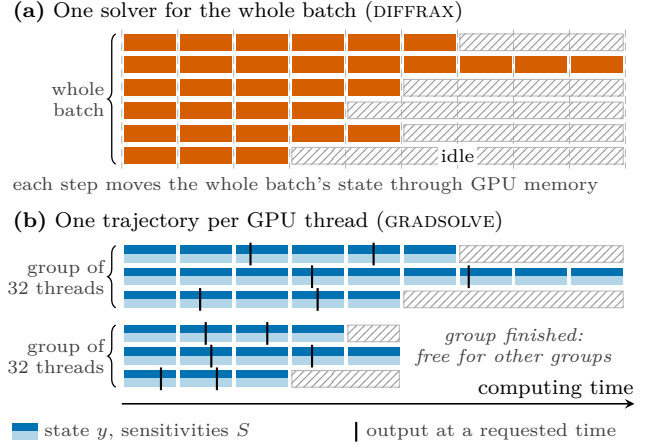
\begin{figure}[t]
\centering
\definecolor{mechred}{HTML}{D55E00}%
\definecolor{mechblue}{HTML}{0072B2}%
\begin{tikzpicture}[x=1cm, y=1cm, font=\footnotesize,
    idle/.style={pattern=north east lines, pattern color=black!40, draw=black!30, line width=0.3pt},
    note/.style={font=\scriptsize, text=black!75, inner sep=0pt},
    brace/.style={decorate, decoration={brace, amplitude=3pt}, line width=0.5pt},
    output/.style={line width=0.8pt, black},
  ]
  \def\mL{1.45}\def\mW{0.74}\def\mH{0.22}\def\mG{0.03}%

  \node[anchor=north west, inner sep=0pt] at (0,0) {\textbf{(a)} One solver for the whole batch (\diffrax)};
  \foreach \mK in {0,...,9}
    \draw[black!30, densely dashed, line width=0.3pt] (\mL+\mK*\mW,-0.38) -- (\mL+\mK*\mW,-2.22);
  \foreach \mN/\mY in {6/-0.55, 9/-0.85, 5/-1.15, 4/-1.45, 5/-1.75, 3/-2.05} {
    \foreach \mK in {1,...,\mN}
      \fill[mechred] (\mL+\mK*\mW-\mW+\mG,\mY-\mH/2) rectangle (\mL+\mK*\mW-\mG,\mY+\mH/2);
    \ifnum\mN<9 \path[idle] (\mL+\mN*\mW+\mG,\mY-\mH/2) rectangle (\mL+9*\mW-\mG,\mY+\mH/2);\fi
  }
  \node[fill=white, inner sep=1pt, font=\scriptsize] at (\mL+6*\mW,-2.05) {idle};
  \draw[brace] (\mL-0.08,-2.05-\mH/2) -- node[left=4pt, note, align=right] {whole\\batch}
    (\mL-0.08,-0.55+\mH/2);
  \node[anchor=north west, note] at (0,-2.33)
    {each step moves the whole batch's state through GPU memory};

  \node[anchor=north west, inner sep=0pt] at (0,-2.78)
    {\textbf{(b)} One trajectory per GPU thread (\gradsolve)};
  \def\mRow#1#2#3#4#5{%
    \foreach \mA/\mB in {#2} {
      \fill[mechblue] (\mL+\mA*\mW+\mG,#1) rectangle (\mL+\mB*\mW-\mG,#1+\mH/2);
      \fill[mechblue!30] (\mL+\mA*\mW+\mG,#1-\mH/2) rectangle (\mL+\mB*\mW-\mG,#1);
    }
    \ifdim #3pt<#4pt \path[idle] (\mL+#3*\mW+\mG,#1-\mH/2) rectangle (\mL+#4*\mW-\mG,#1+\mH/2);\fi
    \foreach \mT in {#5}
      \draw[output] (\mL+\mT*\mW,#1-\mH/2-0.04) -- (\mL+\mT*\mW,#1+\mH/2+0.04);
  }
  \mRow{-3.35}{0/1, 1/2, 2/3, 3/4, 4/5, 5/6}{6}{9}{2.3, 4.5}
  \mRow{-3.65}{0/1, 1/2, 2/3, 3/4, 4/5, 5/6, 6/7, 7/8, 8/9}{9}{9}{3.4, 6.2}
  \mRow{-3.95}{0/1, 1/2, 2/3, 3/4, 4/5}{5}{9}{1.4, 3.5}
  \mRow{-4.40}{0/1, 1/2, 2/3, 3/4}{4}{5}{1.5, 2.6}
  \mRow{-4.70}{0/1, 1/2, 2/3, 3/4, 4/5}{5}{5}{1.6, 3.4}
  \mRow{-5.00}{0/1, 1/2, 2/3}{3}{5}{0.7, 1.7}
  \draw[brace] (\mL-0.08,-3.95-\mH/2) -- node[left=4pt, note, align=right] {group of\\32 threads}
    (\mL-0.08,-3.35+\mH/2);
  \draw[brace] (\mL-0.08,-5.00-\mH/2) -- node[left=4pt, note, align=right] {group of\\32 threads}
    (\mL-0.08,-4.40+\mH/2);
  \node[note, align=center, font=\scriptsize\itshape] at (\mL+7*\mW,-4.58)
    {group finished:\\free for other groups};

  \draw[-stealth, line width=0.6pt] (\mL,-5.33) -- (\mL+9*\mW+0.12,-5.33)
    node[anchor=south east, inner sep=0pt, yshift=2pt] {computing time};
  \fill[mechblue] (0,-5.70) rectangle (0.34,-5.59);
  \fill[mechblue!30] (0,-5.81) rectangle (0.34,-5.70);
  \node[anchor=west, note] at (0.42,-5.70) {state $y$, sensitivities $S$};
  \draw[output] (4.55,-5.83) -- (4.55,-5.57);
  \node[anchor=west, note] at (4.68,-5.70) {output at a requested time};
\end{tikzpicture}
\caption{How the two codes spend GPU time on an ensemble, drawn for six trajectories that need
different numbers of steps. Each box is one step attempt of one trajectory, and computing time runs
to the right. (a) One solver for the whole batch, as in \diffrax{} vectorised with
\code{jax.vmap}: each iteration of the loop (dashed lines) makes one step attempt for every
trajectory, and trajectories that have finished stay idle (hatched) until the slowest one ends. Each step also
reads and writes the state of the whole batch in the GPU's main memory (note under the panel).
(b) One trajectory per GPU thread, as in \gradsolve{}: each trajectory takes its own number of steps,
the dark and light halves of a box are the state and its sensitivities computed in the same step,
and the ticks mark outputs read inside a step. The 32 threads of a group execute together, so a
thread that has finished waits (hatched) only for the slowest thread of its group, and a group that
has finished leaves the GPU to other groups. \citet{SpurioMancini2026gradsolve}
analyses how the difference in cost divides between this waiting and the cost of each step
attempt.}
\label{fig:mechanism}
\end{figure}

\gradsolve{} translates a right-hand side written in \jax{} into one GPU kernel in which each thread
integrates one trajectory with its own step lengths, keeping its state, intermediate evaluations and
sensitivities in memory private to that thread (Figure~\ref{fig:mechanism}, panel b). All three
applications use its explicit kernel, with Tsit5 or Verner~7. Results are returned as \jax{} arrays,
so the likelihood and its derivatives are written in \jax{} around the solve.

\diffrax{}, vectorised over the ensemble with \code{jax.vmap}, instead advances the whole batch
together (Figure~\ref{fig:mechanism}, panel a): every iteration of its loop makes one step attempt
for every trajectory, and the loop runs until the last trajectory has finished, so trajectories
that finish early wait for the slowest. In the \gradsolve{} kernel such waiting is confined to
groups of 32 threads, called warps, which an NVIDIA GPU executes together, so a warp stays occupied
until its slowest member finishes. When all trajectories take the same number of steps the two codes
make the same number of step attempts; the more the step counts differ, the more attempts the batch
loop wastes (equation~\ref{eq:cost}). The cost of one step attempt also differs between the two
codes, and the measured factors combine both effects.

\subsection{How the codes are compared}
\label{sec:comparison}

A tolerance does not fix the accuracy of the solution (Section~\ref{sec:adaptive}), so two codes run
at the same tolerance need not be equally accurate. We therefore compare codes at a fixed accuracy.
For each example we set, before any timing, an accuracy requirement $\epsilon_{\rm req}$ on the
error of each trajectory, in physical units and from a physical argument. For the stellar orbits it
is \stFwdBudget{} of the size and speed of the star's own orbit for the final state, far below the
precision of an observed stream track, and a relative error of \stGradBudget{} for its derivative
(Section~\ref{sec:streams}). For the dark-energy distances it is \deBudget{} of the Hubble distance
$c/H_0$, well below the smallest DESI DR2 distance uncertainty, \deDesiMinErr{}\,$c/H_0$
\citep{DESI2025dr2}. For the precession angles it is \prBudget{}\,rad, the error of the released
\precessionpkg{} package\footnote{\url{https://github.com/dgerosa/precession}} itself (Section~\ref{sec:precession}), so that every accepted code is as
accurate as the code in current use. The error $\epsilon_i$ of trajectory $i$ is measured against a
converged reference, computed with \diffrax{} itself at a tight tolerance
(Appendix~\ref{sec:app-numerics}), and scaled by a physical quantity of that trajectory, never by a value that can
approach zero. A code run at a given tolerance meets the requirement, and we call it accepted, when
\begin{equation}
  P_{99}\bigl(\{\epsilon_i\}_{i=1}^{n}\bigr) \le \epsilon_{\rm req}
  \quad \text{and every } \epsilon_i \text{ is finite},
  \label{eq:accept}
\end{equation}
where $P_{99}$ is the 99th percentile over the $n$ trajectories. The rule bounds the 99th
percentile, so up to one per cent of the trajectories may exceed the requirement; the tables in the
appendix list how many do for each example.

Every code, \gradsolve{} included, runs at the loosest tolerance that meets the requirement on the
timed ensemble; for precession the tolerance is chosen on an independent draw and confirmed, or
tightened, on the timed ensemble (Section~\ref{sec:precession}). For \diffrax{} we try explicit
methods of fifth to eighth order, each at its loosest accepted tolerance and with the ensemble
split into chunks of several sizes, and compare with the fastest accepted configuration, which each
example names. A speed-up factor is the time taken by that configuration divided by the time taken
by \gradsolve{} on the same ensemble. For the stellar-orbit gradient we add a comparison in which
\diffrax{} integrates the same equations with the same method at the tolerance of \gradsolve{}
(Section~\ref{sec:streams}).

All GPU timings use one NVIDIA H200 NVL (141\,GB of memory) in double precision, and
serial codes run on \hwCPU{};
Appendix~\ref{sec:app-numerics} gives the tolerance grid, the references and the timing procedure.

\section{Stellar streams: orbits and their sensitivity to the Galactic potential}
\label{sec:streams}

\begin{figure}[tbp]
\centering
\includegraphics[width=\columnwidth]{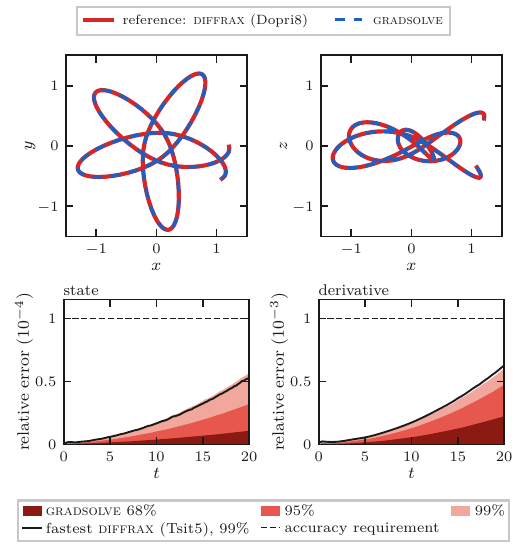}
\caption{Stellar-stream orbits. \emph{Top}: one test orbit in the flattened logarithmic halo,
in units with $v_c = 1$ and $R_c = 0.2$,
projected on the $x$--$y$ and $x$--$z$ planes, from \gradsolve{} (blue, dashed) drawn over the reference,
\diffrax{} (Dopri8) at a tight tolerance (red, solid). \emph{Bottom}: the error against integration time $t$ for the $10^5$ orbits
of the timed ensemble of that size, at equally spaced times up to $t_1$, for the state (left,
equation~\ref{eq:st-fwderr} evaluated at $t$) and for its derivative with respect to $(v_c, q)$
(right, equation~\ref{eq:st-graderr}). Both errors are relative, hence dimensionless. The red bands are the 68th, 95th and 99th percentiles of
the \gradsolve{} error over the orbits at each time. The thin black line is the 99th percentile of
the fastest accepted \diffrax{} configuration, \diffrax{} (Tsit5), which uses the same Runge--Kutta method and tolerance
as \gradsolve{} and, for the derivative, the same sensitivity equations. The dashed line is the
accuracy requirement, which the 99th percentile must meet at $t_1$ (equation~\ref{eq:accept}).}
\label{fig:streams}
\end{figure}

The first application uses all of Section~\ref{sec:method}: an ensemble of orbits together with
their sensitivities. Stars stripped by the tides of the Milky Way from a globular cluster or a dwarf
galaxy spread along nearly the orbit of their progenitor and form a thin stellar stream
\citep[see][for a review]{BonacaPriceWhelan2025review}. Because its track and the velocities along it are set by orbits
in the Galactic potential, a stream probes the mass distribution of the Galaxy \citep[e.g.][]{Johnston1999},
and individual streams have constrained the potential and the flattening of the inner halo
\citep[e.g.][]{Koposov2010gd1, LawMajewski2010, Kupper2015pal5, Bovy2016pal5gd1}. A Fisher-matrix study of eleven known streams found that together they could constrain a simple
analytic Galaxy model to about a per cent \citep{Bonaca2018}, and streams can also map the
acceleration field without an assumed form for the potential \citep[e.g.][]{Nibauer2022}. Modelling a
stream means integrating the orbits of many stars in a trial potential \citep[e.g.][]{Fardal2015},
and such models are increasingly written in \jax{}. For example, \pkg{streamsculptor} \citep{Nibauer2024} integrates its orbits with
\diffrax{} and obtains the response of a stream to perturbations of the potential by forward-mode
differentiation of the equations of motion, the differentiable $N$-body code \pkg{odisseo}
recovers the halo and disc masses, together with the progenitor mass and accretion time, from a
mock GD-1 stream by gradient descent \citep{Viterbo2025odisseo},
and the \pkg{albatross} inference pipeline is built on a \jax{} stream code \citep{Alvey2023albatross}.

Whether the derivatives then drive an optimiser, a sampler or a Fisher forecast, the computation
repeated for every trial potential is many orbits in one potential, each differentiated with
respect to a few parameters that all stars share. We time that computation at one
fixed potential; a full stream fit adds the release of stars from the progenitor and the
likelihood. Each star moves as a test particle in
an axisymmetric flattened logarithmic halo \citep{BinneyTremaine2008},
\begin{equation}
  \begin{gathered}
  \frac{\mathrm{d}^2\mathbf{x}}{\mathrm{d}t^2} = -\nabla\Phi, \\
  \Phi(R, z) = \frac{v_c^2}{2}\,\ln\!\left(R_c^2 + R^2 + \frac{z^2}{q^2}\right),
  \end{gathered}
  \label{eq:loghalo}
\end{equation}
with $R^2 = x^2 + y^2$. Outside the core radius $R_c$ the circular speed approaches $v_c$, and
$q < 1$ flattens the equipotentials. We use units with $v_c = 1$ and take $R_c = 0.2$ and
$q = 0.7$. The state of star $i$ is $y_i = (\mathbf{x}_i, \mathbf{v}_i) \in \mathbb{R}^6$; the halo
parameters are the same for every star, so the stars differ only in their initial conditions,
drawn to span near-circular to nearly radial orbits in and out of the plane. Each orbit conserves
its energy $E_i = \tfrac{1}{2}|\mathbf{v}_i|^2 + \Phi(\mathbf{x}_i)$ and its angular momentum about
the symmetry axis, and the drawn orbits are regular. Every orbit runs from $t_0 = 0$ to a time $t_1$
that spans \stNPeriods{} orbital periods. The forward quantity is the final state $y_i(t_1)$; the
gradient quantity is its sensitivity $S_i(t_1)$ to $v_c$ and $q$, a $6\times2$ matrix obeying
equation~(\ref{eq:sens}).

The energy sets two units for the errors: the zero-velocity radius $r_{E,i}$, at which the potential in
the plane equals the energy and which is the farthest the star can reach, and the speed $v_{E,i}$ the
star would have at the centre,
\begin{equation}
  \Phi(r_{E,i}, 0) = E_i, \qquad \tfrac{1}{2} v_{E,i}^2 + \Phi(0, 0) = E_i .
  \label{eq:st-scales}
\end{equation}
The error of an orbit is the larger of its position and velocity errors at $t_1$ in these units,
\begin{equation}
  \epsilon_i^{y} = \max\!\left( \frac{|\mathbf{x}_i - \mathbf{x}_i^{\rm ref}|}{r_{E,i}},\,
  \frac{|\mathbf{v}_i - \mathbf{v}_i^{\rm ref}|}{v_{E,i}} \right),
  \label{eq:st-fwderr}
\end{equation}
and the error $\epsilon_i^{S}$ of its derivative is the relative error of the whole sensitivity
matrix after its position and velocity rows are divided by $r_{E,i}$ and $v_{E,i}$ and its two
columns multiplied by $v_c$ and $q$, which makes every entry dimensionless
(equation~\ref{eq:st-graderr} in Appendix~\ref{sec:app-numerics}). The accuracy requirement $\epsilon_{\rm req}$ of the acceptance test~(\ref{eq:accept}) is
\stFwdBudget{} for $\epsilon_i^{y}$ and \stGradBudget{} for $\epsilon_i^{S}$
(Section~\ref{sec:comparison}). The reference is
\diffrax{} (Dopri8) integrating the orbit and its sensitivities together at a tight tolerance, with
the error of every component controlled (Appendix~\ref{sec:app-numerics}).

\gradsolve{} integrates one orbit per thread with Tsit5 at its loosest passing tolerance,
\code{rtol} = \code{atol} = \stRtol{}; for the gradient it integrates the sensitivity equations in
the same launch, with the step size controlled on the orbit alone. The forward candidates are
\diffrax{} (Tsit5) and \diffrax{} (Dopri8), applied to the batch. For the gradient \diffrax{} offers two
formulations: \code{jax.jacfwd} through the discretised solve, and equation~(\ref{eq:sens})
integrated as one larger ODE with the step size again controlled on the orbit. Because \diffrax{}
holds its step sizes fixed when it differentiates, both return the derivative of the discrete solution
(Section~\ref{sec:sensitivities}) and give bit-identical Jacobians; both are candidates for the
fastest accepted configuration.

We report two comparisons for the gradient. The first is against the fastest accepted \diffrax{}
configuration, the factor a user who picks the fastest set-up would see. The second
holds the equations, the Runge--Kutta method (Tsit5) and the tolerance fixed: \diffrax{} integrates
equation~(\ref{eq:sens}) at the tolerance of \gradsolve{} rather than at its own loosest passing
one. The step-size controllers still differ (equation~\ref{eq:controller} and the text after it).
At this tolerance \diffrax{} makes slightly more
step attempts, a mean $N_i$ of \stStepsDfx{} against \stStepsGs{} for \gradsolve{}.
Figure~\ref{fig:speedup} (left) shows the result, and Table~\ref{tab:streams} in
Appendix~\ref{sec:app-tables-sp} gives the full timings. \gradsolve{} obtains the
orbits and their derivatives between \stGradFactorDrawnEfour{} and \stGradFactorDrawnEfive{} times
faster than the fastest accepted \diffrax{} configuration, and the orbits alone between
\stFwdFactorDrawnEsix{} and \stFwdFactorDrawnEfive{} times faster; both factors are largest at
$n=10^5$. At $n=10^5$ and $10^6$ the fastest accepted configuration for the derivative is
\diffrax{} integrating the sensitivity equations with the same Runge--Kutta method and tolerance as
\gradsolve{}, so the two comparisons give the same factor. At $n=10^4$ it is \stGradDfxArmEfour{},
and the comparison at fixed method gives \stGradSameFactorDrawnEfour{}.
On one CPU core, \gala{}\footnote{\url{https://github.com/adrn/gala}} \mbox{(DOPRI853)} \citep{PriceWhelan2017gala}, integrating one star at a time
with this eighth-order Dormand--Prince method at its loosest passing tolerance and differentiating by central
finite differences, takes \stCpuUs{}\,$\mu$s per orbit for the orbit and its derivatives. Run on all 128 cores of the
CPU, one process per core, it is \cpuNodeSpeedupLo{} to \cpuNodeSpeedupHi{} times faster than on one core, and
\gradsolve{} is still \stNodeSpeedupGsEsix{} times faster than the whole CPU at $n=10^6$, while the fastest
\diffrax{} configuration is slower than it (Figure~\ref{fig:speedup}). For the orbits and their
derivatives at $n=10^6$, the choice of GPU code thus decides whether moving from the CPU to the GPU
pays off at all.

Over the $10^6$ timed orbits the 99th percentile of the \gradsolve{} error is \stFwdErrPnn{} for
the final state and \stGradErrPnn{} for its derivative, within the accuracy requirement.
Figure~\ref{fig:streams} follows the error along the integration for the $10^5$ orbits of the
smaller timed ensemble: it grows with $t$ in both codes, and the 99th percentile of each stays
below the requirement at every output time.

In a stream model these would be the orbits of the released stars, and a gradient-based fit needs
$y_i(t_1)$ and $S_i(t_1)$ at every trial potential, as a Fisher forecast such as that of
\citet{Bonaca2018} does at the fiducial one. A Galaxy model with disc and bulge as well as halo
parameters differentiates more of them; each one adds six components per star, so the cost of the
gradient grows with $m$, and for many parameters the reverse-mode gradient applies
(Section~\ref{sec:discussion}).

\section{Dark energy: distances for many cosmologies and a profile likelihood}
\label{sec:darkenergy}

\begin{figure}[tbp]
\centering
\includegraphics[width=\columnwidth]{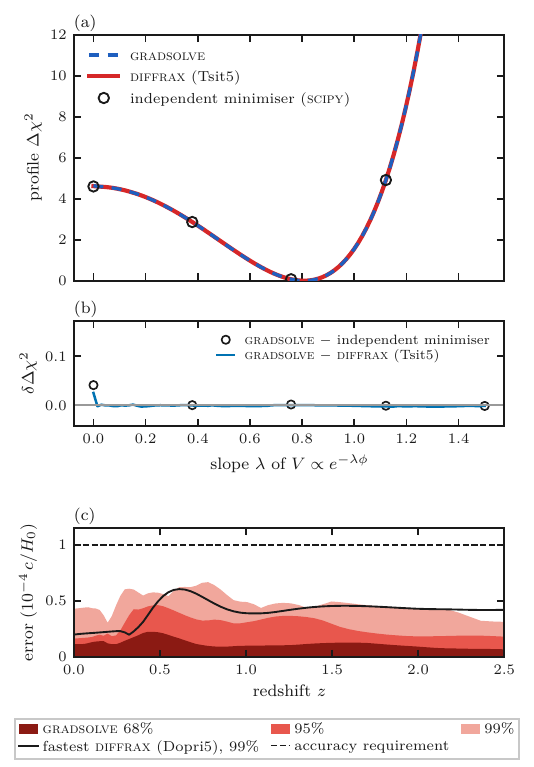}
\caption{Dark energy. (a) The profile $\Delta\chi^2(\lambda)$ on the observed DESI DR2 and
Pantheon+ data from \gradsolve{}, blue and dashed, drawn over \diffrax{} (Tsit5), red and solid,
with the independent minimiser (\scipy{}) at selected values of $\lambda$ as circles.
(b) The difference in $\Delta\chi^2$ between \gradsolve{} and \diffrax{} (Tsit5), as a line, and
between \gradsolve{} and the independent minimiser, as circles. (c) The distance error against redshift $z$
for $10^5$ cosmologies with the amplitude supplied, at \deNzDense{} output redshifts spaced
uniformly in $\ln(1+z)$; the error at each redshift is the larger of $|\Delta D_M|$ and
$|\Delta D_H|$, in units of $c/H_0$. The red bands are the 68th, 95th and 99th percentiles of the
\gradsolve{} error over the cosmologies at each redshift, and the thin black line is the 99th
percentile of the fastest accepted \diffrax{} configuration, \diffrax{} (Dopri5), each code at the tolerance at which it
was timed. The dashed line is the accuracy requirement. It applies to the 99th percentile of the
largest error of each cosmology over redshift (equation~\ref{eq:accept}), which is at least as
large as the 99th percentile at any single redshift.}
\label{fig:darkenergy}
\end{figure}

Where the stream example differentiates many orbits at one potential, the second application
first scans the expansion history over many cosmologies and then differentiates it inside an
iterative fit. Baryon acoustic oscillations (BAO) measured in the second DESI data release, combined with the
cosmic microwave background and type Ia supernovae, prefer a dark-energy equation of state that
departs from the cosmological-constant value $w=-1$ at late times \citep{DESI2025dr2} when $w$ is
parametrised as $w(a) = w_0 + w_a(1-a)$ \citep{Chevallier2001cpl, Linder2003cpl}. A physical model
with this behaviour is thawing quintessence \citep{CaldwellLinder2005, ScherrerSen2008}, a light
scalar field \citep{RatraPeebles1988, Wetterich1988, Caldwell1998quintessence}; see
\citet{Copeland2006} for a review. The field is held at rest by the expansion in the early universe,
so that $w \simeq -1$, and starts to roll down its potential recently, so that $w$ rises above $-1$.
Whether the data favour such a field over a cosmological constant is debated
\citep[e.g.][]{Efstathiou2025, Wolf2024, Lodha2025desi}. We take the exponential potential $V(\phi) =
V_0\,\mathrm{e}^{-\lambda\phi}$ of \citet{Wetterich1988}, used in recent fits to the DESI data
\citep[e.g.][]{Ramadan2024}, for which $\lambda = 0$ recovers a cosmological constant.

The background of one such cosmology takes \deFwdScipyCond{}\,$\mu$s to integrate with \scipy{}\footnote{\url{https://github.com/scipy/scipy}} (LSODA)
\citep{Virtanen2020scipy} on one CPU core, but an analysis repeats it many times: at every point a sampler visits, for every cosmology in a forecast
or an emulator training set \citep[e.g.][]{SpurioMancini2022cosmopower, Nygaard2023connect,
CosmoPowerJAX2023}, and, in a profile likelihood whose coverage
is tested on simulated data, at every iteration of every minimisation on every mock data set.
After setting out the equations we time two such workloads: the distances for a population of
cosmologies, and the minimisations of a profile likelihood.

\subsection{Background equations and spatial flatness}
\label{sec:de-model}

We integrate in the number of e-folds $N = \ln a$, which plays the role of $t$ in Section~\ref{sec:method} and is unrelated to the step count $N_i$, with $N=0$ today and
$N_z = -\ln(1+z)$ at redshift $z$. The field is in reduced Planck units, densities are in units of
the critical density today, and $E = H/H_0$, so that the potential becomes
$v(\phi) = a_0\,\mathrm{e}^{-\lambda\phi}$ with a dimensionless amplitude $a_0$. For one cosmology
the state is $y = (\phi, u, I)$: the field, its velocity $u = \mathrm{d}\phi/\mathrm{d}N$, and a
quantity $I$ that accumulates the comoving distance. The parameters are
$\theta = (a_0, \lambda, \Omega_{\rm m}, \Omega_{\rm r})$, with the radiation density fixed. The
Friedmann equation gives the expansion rate from the state, and the three variables evolve as
\begin{equation}
\begin{gathered}
  E^2\Bigl(1 - \frac{u^2}{6}\Bigr) = \Omega_{\rm m}\mathrm{e}^{-3N} + \Omega_{\rm r}\mathrm{e}^{-4N}
    + v(\phi), \\
  \frac{\mathrm{d}\phi}{\mathrm{d}N} = u, \qquad
  \frac{\mathrm{d}u}{\mathrm{d}N} = F(\phi, u, N; \theta), \qquad
  \frac{\mathrm{d}I}{\mathrm{d}N} = \frac{\mathrm{e}^{-N}}{E},
\end{gathered}
\label{eq:de-model}
\end{equation}
where $F$ is the Klein--Gordon equation of the field written in e-folds
(Appendix~\ref{sec:app-de}, equation~\ref{eq:de-rhs}). The field starts at rest, $\phi = u = I = 0$,
at redshift \deZstart{}. The distances measured by BAO follow from the solution as
$D_M(z) = (c/H_0)\,[I(0) - I(N_z)]$ and $D_H(z) = c/[H_0\,E(N_z)]$, so $H_0$ does not enter
equation~(\ref{eq:de-model}) and only converts the dimensionless solution into distances.

The amplitude $a_0$ is not a free parameter, because a spatially flat model must have $E(0) = 1$:
the densities today add up to the critical density. With $\phi_0$ and $u_0$ the field and its
velocity at $N=0$, this condition reads
\begin{equation}
  g(a_0; \lambda, \Omega_{\rm m}) \equiv E(0)^2 - 1
  = \frac{\Omega_{\rm m} + \Omega_{\rm r} + a_0\,\mathrm{e}^{-\lambda\phi_0}}{1 - u_0^2/6} - 1 = 0,
  \label{eq:de-flat}
\end{equation}
and it fixes $a_0(\lambda, \Omega_{\rm m})$. Because $\phi_0$ and $u_0$ are known only after the
integration to today, we solve equation~(\ref{eq:de-flat}) by Newton's method in $a_0$, which is the
shooting method for a boundary-value problem with one unknown. Each Newton step is one integration
to $N=0$ that also carries the sensitivity $S_{a_0} = \partial y/\partial a_0$, from which follows
the slope $\mathrm{d}g/\mathrm{d}a_0$. The iteration starts from the value
$a_0 = 1 - \Omega_{\rm m} - \Omega_{\rm r}$, which would be exact if the field never moved. Used
without solving for $a_0$, this value leaves $E(0)$ below one (Appendix~\ref{sec:app-de}): at the
largest slope of the profile grid $E(0) = \deFrozenEzero{}$, and on the observed data the
best-fitting $\Omega_{\rm m}$ moves by up to \deFrozenOmShift{} standard deviations of the fit with
the flat model.

A fit that varies $\Omega_{\rm m}$ also needs the derivative of $a_0$ with respect to
$\Omega_{\rm m}$, because the observables depend on $\Omega_{\rm m}$ both directly and through the
amplitude that keeps the model flat. Differentiating
$g\bigl(a_0(\lambda, \Omega_{\rm m}); \lambda, \Omega_{\rm m}\bigr) = 0$ gives, by the
implicit-function theorem,
\begin{equation}
  \frac{\mathrm{d}a_0}{\mathrm{d}\Omega_{\rm m}} = -\frac{\mathrm{d}g/\mathrm{d}\Omega_{\rm m}}
    {\mathrm{d}g/\mathrm{d}a_0}, \qquad
  \frac{\mathrm{d}g}{\mathrm{d}\theta_j} = \frac{\partial g}{\partial y}\,S_{\theta_j}(0)
    + \frac{\partial g}{\partial\theta_j},
  \label{eq:de-ift}
\end{equation}
where each total derivative combines the explicit dependence of $g$ on $\theta_j$ with its
dependence through the state today, carried by the sensitivity $S_{\theta_j} = \partial y/\partial\theta_j$
at $N=0$. The derivative of the state along the flat models is then
$\mathrm{d}y/\mathrm{d}\Omega_{\rm m} = S_{\Omega_{\rm m}} + (\mathrm{d}a_0/\mathrm{d}\Omega_{\rm m})\,S_{a_0}$.
\gradsolve{} integrates both sensitivities in the same GPU launch as the solution, so once $a_0$ is
known the derivative through the flatness condition requires no further integration.

\subsection{Distances for a population of cosmologies}
\label{sec:de-forward}

We compute the distances at the \deNz{} effective redshifts of the DESI DR2 BAO measurements for
$10^4$ to $10^6$ cosmologies, with $\lambda$ and $\Omega_{\rm m}$ drawn uniformly from
$\dePriorLam{}$ and $\dePriorOm{}$. The error $\epsilon_i$ of cosmology $i$ is the largest
difference in $D_M$ or $D_H$ over the output redshifts from a converged reference, \diffrax{} (Tsit5) at
a tight tolerance, in units of the
Hubble distance $c/H_0$. The accuracy requirement is $\epsilon_{\rm req} = \deBudget{}$ in these units (Section~\ref{sec:comparison}).
The \gradsolve{} arm uses the explicit kernel with Tsit5 and reads the distances from its dense
output, and the \diffrax{} arm is the fastest accepted \diffrax{} configuration. For
each arm we report two costs: the integration alone for a given amplitude (\emph{amplitude
supplied}), and the full cost from $(\lambda, \Omega_{\rm m})$ to the distances, which adds the
arm's own Newton steps on equation~(\ref{eq:de-flat}) (\emph{including the solve for $a_0$}). On the CPU,
\scipy{} (LSODA) \citep{Hindmarsh1983odepack, Petzold1983lsoda} integrates one cosmology at a time on
one core.

Figure~\ref{fig:speedup} shows the factor over \diffrax{} against the number of cosmologies, and
Table~\ref{tab:de_forward} in Appendix~\ref{sec:app-de} gives the costs. For a million cosmologies
\diffrax{} costs \deFwdFactorCondEsix{} times as much as \gradsolve{} with the amplitude supplied
and \deFwdFactorFullEsix{} times as much including the solve for $a_0$. With the amplitude supplied,
\gradsolve{} then takes \deFwdGsCondEsix{}\,$\mu$s per cosmology, against \deFwdScipyCond{}\,$\mu$s
for \scipy{} (LSODA) on one CPU core; with \scipy{} run on all 128 cores, one process per core,
\gradsolve{} is still \deNodeSpeedupGsEsix{} times faster than the whole CPU (Figure~\ref{fig:speedup}).
Over the $10^6$ cosmologies the 99th
percentile of the \gradsolve{} distance error is \deErrPnn{}, within the accuracy requirement.
Figure~\ref{fig:darkenergy}(c) shows how the error depends on redshift, for $10^5$ cosmologies on
a finer grid of \deNzDense{} output redshifts, with both codes at the tolerances at which they were
timed. For a million cosmologies \gradsolve{} is thus faster than both \diffrax{} and the whole CPU
while meeting the accuracy requirement.

\subsection{A profile likelihood}
\label{sec:de-profile}

The profile likelihood of the slope $\lambda$ asks how well the data can be fitted at each fixed
value of $\lambda$. We fix $\lambda$ at each of \deProfGrid{} values on a grid, find the
$\Omega_{\rm m}$ and $H_0$ that minimise $\chi^2$ at that value, and record the minimum; the
profile $\Delta\chi^2(\lambda)$ is this minimum measured from its lowest value along the grid. The
data are the DESI DR2 BAO distances \citep{DESI2025dr2} and the Pantheon+ supernova magnitudes
\citep{Scolnic2022pantheon, Brout2022}, and $\chi^2$ compares the model with both through their
published covariances (Appendix~\ref{sec:app-de} gives the data model).

Each minimisation is a Gauss--Newton fit. At every iteration it needs the residuals between model
and data and their derivatives with respect to $\Omega_{\rm m}$ and $H_0$, and from these it
computes the step towards the minimum. The derivative with respect to $H_0$ is analytic, because
$H_0$ only rescales the distances and the sound horizon. The derivative with respect to
$\Omega_{\rm m}$ must follow the model along flat cosmologies, so it needs
$\mathrm{d}y/\mathrm{d}\Omega_{\rm m}$ from equation~(\ref{eq:de-ift}) at every output redshift.
\gradsolve{} integrates $S_{a_0}$ and $S_{\Omega_{\rm m}}$ in the same launch as the solution, so a
single launch per iteration returns the residuals and all their derivatives, including the
derivative through the flatness condition. The \diffrax{} (Tsit5) arm obtains the same derivatives by
forward-mode differentiation of its batched solve, and \scipy{} (LSODA) by
finite differences of its solutions on one CPU core; all three apply the same update to the same
residuals. With the solve for $a_0$ included, each iteration first takes a few Newton steps on
equation~(\ref{eq:de-flat}) for the new $\Omega_{\rm m}$.

One profile on the observed data takes \deProfGrid{} minimisations. Testing whether the resulting
interval has the stated coverage means repeating the profile on many mock data sets, drawn from a
fiducial model with Gaussian noise from the published covariances, and this is the workload we
time: \deProfMocks{} mock data sets, or \deProfN{} minimisations run together on one GPU. Each minimisation runs a fixed \deProfIters{} iterations, and every one must then pass a convergence test (\deProfConv{}).

Figure~\ref{fig:darkenergy} shows the profile on the observed data. It is smallest at
$\lambda \simeq \deObsLamBest{}$, and a cosmological constant ($\lambda = 0$) lies
\deObsDchiZero{} higher in $\chi^2$. The profiles from \gradsolve{} and \diffrax{} (Tsit5) differ by at
most \deObsArmsMax{} in $\chi^2$, and both agree with an independent minimiser (\scipy{}) run on the
CPU to within \deObsIndepMax{}. These differences are far smaller than the \deObsDchiZero{} in
$\chi^2$ separating $\lambda = 0$ from the best fit, so the choice of code does not change the result.

With the solve for $a_0$ included, one iteration costs \deProfGsFull{}\,$\mu$s per minimisation with
\gradsolve{}, \deProfDfxFull{}\,$\mu$s with \diffrax{} and \deProfScipyFull{}\,$\mu$s with
\scipy{} (LSODA), so the factor over \diffrax{} is \deProfFactorFull{} for the whole
iteration and \deProfSolveFactorFull{} for the differentiated solve alone, the rest of the
iteration being common to both GPU codes (Table~\ref{tab:de_profile} in Appendix~\ref{sec:app-de}).

\section{Spin precession across a black-hole population}
\label{sec:precession}

\begin{figure}[tbp]
\centering
\includegraphics[width=\columnwidth]{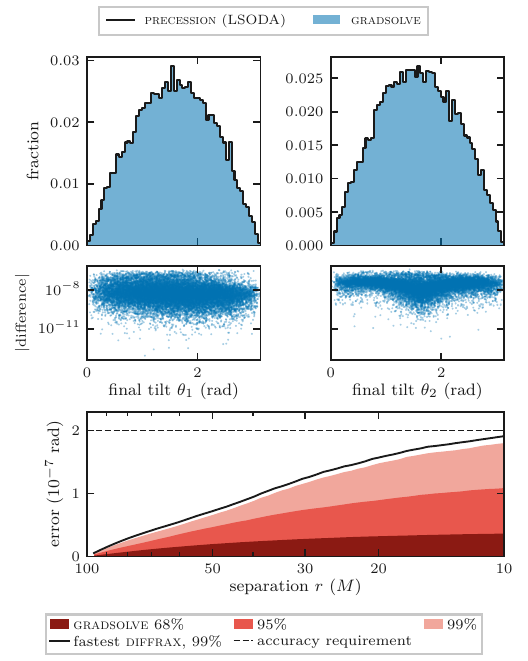}
\caption{Spin precession. \emph{Top}: distributions of the two final spin tilts, the angles between
each spin and $\hat{\mathbf{L}}$ at $r=10\,M$, from \precessionpkg{} (LSODA) and from
\gradsolve{} (Verner~7), on the same \prParityN{} binaries, with the absolute per-binary difference
between the two codes below each. \emph{Bottom}: on the same binaries, the error $\epsilon_i$
(equation~\ref{eq:prec-err}, evaluated at each separation) as a function of the separation $r$, from $100\,M$ on the left to $10\,M$ on the right. The red bands are the 68th,
95th and 99th percentiles of the \gradsolve{} error over the binaries at each separation,
and the thin black line is the 99th percentile of the fastest accepted \diffrax{} configuration at
$n=10^6$, \prDfxArm{}; both codes run at their timed tolerance. The dashed line is the accuracy
requirement, which the 99th percentile must meet at $r=10\,M$ (equation~\ref{eq:accept}).}
\label{fig:precession}
\end{figure}

The spins of merging black holes carry a record of how the binaries formed. Their orientations
relative to the orbit depend on the formation channel \citep[e.g.][]{Rodriguez2016spins,
Vitale2017spins} and on individual stages of massive-binary evolution, such as the kicks the black
holes receive at birth \citep[e.g.][]{Kalogera2000}, so measured spin orientations can constrain those
stages \citep[e.g.][]{Gerosa2018spinorientations}, and gravitational-wave data can distinguish
populations with spins aligned to the orbit from isotropic ones \citep[e.g.][]{Farr2017spins}.
The spins are set when the black holes
form but are measured near merger, and in between the spins and the orbital angular momentum
precess about one another while gravitational-wave emission shrinks the orbit. A formation model
is therefore compared with data by evolving every binary of a synthetic catalogue from the
separation at which it forms down to the detector band \citep[e.g.][]{Gerosa2018spinorientations}, and
inference on the model repeats this for every binary of the catalogue at every trial set of its
parameters. This third application uses the forward integration alone, with a state of nine
components and an accuracy requirement set by the error of \precessionpkg{}; we time populations of $10^4$ to
$10^6$ binaries.

The evolution is organised by three widely separated time-scales \citep[e.g.][]{Kesden2015, GerosaKesden2016precession}:
at separation $r$, measured in gravitational radii, the orbital period grows as $r^{3/2}$, the
precession period as $r^{5/2}$ and the time on which radiation reaction shrinks the orbit as $r^{4}$.
\precessionpkg{} \citep{GerosaKesden2016precession,
Gerosa2023precessionv2} uses this ordering in two ways. Its orbit-averaged evolution averages over
the orbit and follows every precession cycle; its precession-averaged evolution also averages over
the cycles \citep{Kesden2015} and can start from arbitrarily large separations. We integrate the orbit-averaged
equations, which \precessionpkg{} solves one binary at a time with LSODA \citep{Hindmarsh1983odepack,
Petzold1983lsoda} through \scipy{} \citep{Virtanen2020scipy}.

We use units with $G=c=M=1$, where $M=m_1+m_2$ is the total mass. A binary has mass ratio
$q=m_2/m_1\le1$, symmetric mass ratio $\eta=q/(1+q)^2$, dimensionless spins $\chi_j$ and spin
magnitudes $S_j=\chi_j m_j^2$ ($j=1,2$; unrelated to the sensitivities $S_i$ of
Section~\ref{sec:sensitivities}), and orbital velocity $v=r^{-1/2}$; at Newtonian order the
orbital angular momentum has magnitude $L=\eta/v$. The orbit-averaged equations evolve the unit
vectors $\hat{\mathbf{L}}$, $\hat{\mathbf{S}}_1$ and $\hat{\mathbf{S}}_2$
\citep{Apostolatos1994, Kidder1995, GerosaKesden2016precession}:
\begin{align}
\frac{\mathrm{d}\hat{\mathbf{S}}_j}{\mathrm{d}t} &= \boldsymbol{\Omega}_j\times\hat{\mathbf{S}}_j ,
\label{eq:prec-spins}\\
\frac{\mathrm{d}\hat{\mathbf{L}}}{\mathrm{d}t} &= -\frac{v}{\eta}\sum_{j=1}^{2}
S_j\,\frac{\mathrm{d}\hat{\mathbf{S}}_j}{\mathrm{d}t} ,
\label{eq:prec-L}
\end{align}
with the precession frequency of the first spin
\begin{equation}
\boldsymbol{\Omega}_1 = \frac{v^6}{2}\Big[(4+3q)\,\mathbf{L} + \mathbf{S}_2
- 3\,\hat{\mathbf{L}}\cdot\big(\mathbf{S}_2 + q\,\mathbf{S}_1\big)\,\hat{\mathbf{L}}\Big] ,
\label{eq:prec-omega}
\end{equation}
and $\boldsymbol{\Omega}_2$ obtained by exchanging the labels 1 and 2 and replacing $q$ by $1/q$. The
term in $\mathbf{L}$ is the spin--orbit coupling and the others are spin--spin couplings.
Equation~(\ref{eq:prec-L}) keeps the total angular momentum $\mathbf{L}+\mathbf{S}_1+\mathbf{S}_2$
fixed on the precession time-scale. The magnitude $L$ changes only through radiation reaction,
\begin{equation}
\frac{\mathrm{d}v}{\mathrm{d}t} = \frac{32}{5}\,\eta\,v^{9}\Big(1+\sum_{k=2}^{7} b_k\,v^k\Big) ,
\label{eq:prec-rr}
\end{equation}
where the post-Newtonian coefficients $b_k$, as implemented in \precessionpkg{}, depend on the masses,
the spins and the angles between the three unit vectors, and $b_6$ also on $\ln v$. Like
\precessionpkg{}, we integrate in $v$, which grows monotonically during the inspiral. In the notation of
Section~\ref{sec:method}, the state is $y=(\hat{\mathbf{L}},\hat{\mathbf{S}}_1,\hat{\mathbf{S}}_2)
\in\mathbb{R}^9$, the independent variable is $v$ in place of $t$, the right-hand side is
equations~(\ref{eq:prec-spins}) and (\ref{eq:prec-L}) divided by $\mathrm{d}v/\mathrm{d}t$, the
parameters are $\theta=(q,\chi_1,\chi_2)$, and every binary runs from $r=100\,M$ to $10\,M$. Our \jax{}
right-hand side reproduces that of \precessionpkg{} at random states. Binaries are drawn with $q$ uniform
in $[0.1,1]$, $\chi_j$ uniform in $[0,1]$ and isotropic spin directions at $r=100\,M$, and the
initial vectors are built with the conversion from angles in \precessionpkg{}, so every arm starts from
the same states.

Equations~(\ref{eq:prec-spins}) and (\ref{eq:prec-L}) rotate the three unit vectors without changing
their lengths, and after orbit averaging the shortest time-scale left is the precession period,
which the solution itself follows. Steps chosen for accuracy are therefore stable, and we use the
explicit kernel. Adaptive step control still matters: at leading order the number of precession cycles per
unit $\ln r$ scales as the ratio of the radiation-reaction time to the precession period,
$\propto r^{3/2}$, so most cycles occur near the starting separation and the solver can lengthen its
steps as the binary tightens. The number of steps also differs from binary to binary, and a batch that advances all binaries
together waits for the slowest (Section~\ref{sec:library}).

Every arm, \precessionpkg{} included, is judged against the same converged reference,
\diffrax{} (Dopri8) at a tight tolerance (Appendix~\ref{sec:supp-reference}). The error of binary $i$ is the largest angle
between the final directions from the arm and from the reference,
\begin{equation}
\epsilon_i = \max_{\hat{\mathbf{X}}\in\{\hat{\mathbf{L}},\hat{\mathbf{S}}_1,\hat{\mathbf{S}}_2\}}
\angle\big(\hat{\mathbf{X}}_i,\hat{\mathbf{X}}_i^{\rm ref}\big)\Big|_{r=10\,M} .
\label{eq:prec-err}
\end{equation}
Unlike the error in the azimuthal angle between the two spins, which diverges as either spin
approaches alignment with $\hat{\mathbf{L}}$, $\epsilon_i$ does not depend on the choice of
coordinates. We set the accuracy requirement from \precessionpkg{} itself: over \prParityN{} binaries its 99th
percentile of $\epsilon_i$ is \prRelErrPnn{}\,rad, and rounding it up to one significant figure
gives $\epsilon_{\rm req}=\prBudget{}$\,rad. An arm is accepted when 99 per cent of the binaries are
within $\epsilon_{\rm req}$ (Section~\ref{sec:comparison}), so every GPU arm matches the accuracy of
\precessionpkg{} at the 99th percentile.

\gradsolve{} runs one binary per thread with Tsit5 or Verner~7, each at the loosest tolerance on the
grid that passes: \code{rtol} = \code{atol} = \prTsitRtol{} for Tsit5 and \prVernRtol{} for
Verner~7. At the looser \prTsitOldRtol{} the 99th percentile of Tsit5 on the draw used to select
tolerances is \prTsitOldErrPnn{}\,rad, above the requirement, so the tighter value is the one timed. For
\diffrax{} we try Tsit5, Verner~7 and Dopri8, each at its own loosest passing tolerance and with the
batch split into chunks in several ways. \diffrax{} has no Verner method, so its Verner~7
arm is a \diffrax{} explicit Runge--Kutta solver built from the same coefficients as the \gradsolve{}
kernel. The fastest accepted configuration is \prDfxArm{}. For
scale, \precessionpkg{} takes \prCpuMs{}\,ms per binary on one core of the \cpuModel{} host.

Figure~\ref{fig:speedup} (right) shows the result, and Table~\ref{tab:precession} in
Appendix~\ref{sec:app-tables-sp} gives the full timings. For a population of \prN{} binaries
\gradsolve{} (Verner~7) is \prFactorVernDrawn{} times faster than the fastest accepted \diffrax{}
configuration, and \gradsolve{} (Tsit5) \prFactorTsitDrawn{} times faster. \gradsolve{} (Verner~7)
takes \prVernDrawn{}\,$\mu$s per binary, against \prCpuMs{}\,ms for \precessionpkg{} (LSODA) on one
CPU core; with \precessionpkg{} run on all 128 cores, one process per core, \gradsolve{} is still
\prNodeSpeedupGsEsix{} times faster than the whole CPU (Figure~\ref{fig:speedup}). The factor rises from
$n=10^4$ to $10^5$ and then levels off. \diffrax{} (Verner~7) and \gradsolve{} (Verner~7)
make almost the same number of step attempts, a mean $N_i$ of \prStepsDfx{} against
\prStepsGs{}.
Over the \prN{} timed binaries the 99th percentile of $\epsilon_i$ is \prVernErrPnn{}\,rad for
\gradsolve{} (Verner~7) and \prTsitErrPnn{}\,rad for \gradsolve{} (Tsit5). Both \gradsolve{} methods
meet the accuracy requirement, and since both Verner~7 codes make almost equal numbers of step
attempts, the factor of \prFactorVernDrawn{} comes from the cost per attempt.

Figure~\ref{fig:precession} compares the final spin tilts, the angles between each spin and
$\hat{\mathbf{L}}$, from \gradsolve{} and from \precessionpkg{} on the same \prParityN{}
binaries; the per-binary difference between the two contains the error of both codes. Its lower
panel follows the error of \gradsolve{} and \diffrax{} against the converged reference along the
inspiral of the same binaries. For \diffrax{} the values between steps are computed as in
\gradsolve{} (Verner~7), by a shortened step from the start of the accepted step
(Appendix~\ref{sec:app-numerics}), because its built-in interpolation for this method is much less
accurate than the integration itself. The error grows as the binaries tighten, and at $r=10\,M$ the
99th percentile of both codes is within the accuracy requirement.

\section{Discussion}
\label{sec:discussion}

The three applications share the conditions for which the per-thread kernel was built: a
closed-form right-hand side, a state of three to nine components, a non-stiff problem, at most two
differentiated parameters, and ensembles of $10^4$ to $10^6$ members. Under these conditions
\gradsolve{} is faster than the fastest accepted \diffrax{} configuration on the same GPU in every
configuration we timed, for the integration alone and for the derivatives
(Figure~\ref{fig:speedup}), but the factor varies by about an order of
magnitude between examples and ensemble sizes. We first ask what the comparisons at fixed method say
about where the factor comes from, then what sets its size, and finally where the approach applies and
where other methods suit better.

The stellar-orbit gradient separates implementation from method. The two \diffrax{} formulations
are one method written two ways (Section~\ref{sec:streams}), and \gradsolve{} integrates the same
equations with the same Runge--Kutta method. At
$n=10^5$ and $10^6$ the fastest accepted \diffrax{} configuration is exactly this
computation, \stGradDfxArm{}, at the tolerance \gradsolve{} uses, so the factor a user would see
and the factor at fixed method and tolerance are the same number (Table~\ref{tab:streams} in Appendix~\ref{sec:app-tables-sp}). Since
the two codes also make almost the same number of step attempts (\stStepsDfx{} against
\stStepsGs{} per orbit), although their step-size controllers differ, the factor per step attempt is
essentially the factor in time: the gap lies in the cost
of one step attempt, a property of the implementation rather than of the numerical method. Both
arms pass the same accuracy test, and their tails differ in both directions: at $n=10^6$,
\stGradNover{} orbits exceed the accuracy requirement for the derivative with \gradsolve{} against
\stGradDfxNover{} with \diffrax{}, and \stFwdNover{} against \stFwdDfxNover{} exceed it for the final
state, all far below
the one per cent of orbits that the acceptance test allows.
In the precession example the fastest accepted \diffrax{} configuration at $n=\prN{}$, \prDfxArm{}, is built
from the same coefficients as the faster \gradsolve{} arm, runs at the same tolerance and makes
almost the same number of step attempts (\prStepsDfx{} against \prStepsGs{} per binary), so its
factor of \prFactorVernDrawn{} again compares two implementations of one method, up to the
step-size controller.

Two differences in implementation raise the cost per step attempt, the total time divided by
$\sum_i N_i$. A batch that advances all trajectories together keeps
every trajectory in the loop until the slowest has finished, the waiting modelled by
equation~(\ref{eq:cost}); and a code that works on whole arrays, as \diffrax{} under \code{jax.vmap} does, typically writes the state of the whole batch to the GPU's main memory and reads it back between the operations of a step, whereas the per-thread kernel keeps each
trajectory's state, stages and sensitivities in memory private to its thread
(Section~\ref{sec:library}). For reverse-mode gradients \citet{SpurioMancini2026gradsolve} found
that uneven step counts contribute only a small factor, and traced most of the gap to the form of the
differentiated computation: a fixed-length replay of the recorded steps, which the GPU executes
faster than an adaptive loop differentiated in place.

The dependence on ensemble size in Figure~\ref{fig:speedup} is consistent with costs that are fixed per call, such as
launching the computation: with few trajectories they weigh on every trajectory, and the cheaper
each integration is, the larger the ensemble must be before they become negligible. For the stellar
orbits the \gradsolve{} cost per orbit falls from \stFwdGsDrawnEfour{} to
\stFwdGsDrawnEfive{}\,$\mu$s between $n=10^4$ and $10^5$ and then barely changes, while the
\diffrax{} cost keeps falling up to $n=10^6$; the stellar-orbit factors therefore peak at $n=10^5$
and are smaller at $n=10^6$, \stFwdFactorDrawnEsix{} for the orbits and \stGradFactorDrawnEsix{} for
the gradient. In the precession example both costs per binary stop falling beyond $n=10^5$ and the
factor levels off (Figure~\ref{fig:speedup}). The dark-energy trajectories are the cheapest
of the three, and the \gradsolve{} cost per cosmology is still falling between $n=10^5$ and $10^6$, from
\deFwdGsCondEfive{} to \deFwdGsCondEsix{}\,$\mu$s; the dark-energy factors grow with ensemble
size over the whole range.

Once the solve is fast, the rest of the analysis sets the pace. In the dark-energy profile
likelihood the differentiated solve, including the solve for $a_0$, becomes \deProfSolveFactorFull{}
times cheaper with \gradsolve{} and the whole Gauss--Newton iteration \deProfFactorFull{} times
cheaper; the solve then costs less than whitening and projecting the residuals
(Table~\ref{tab:de_profile}), so the differential equation is no longer the bottleneck of the fit.

The number of differentiated parameters, the size of the state and the stiffness of the equations
decide where the approach applies. The fits and forecasts timed here need the derivative of every
residual, so forward sensitivities are the cheaper choice (equation~\ref{eq:disc-modes}); for a
sampler that needs only the gradient of the log-likelihood with respect to many parameters,
reverse mode is. Lens models with thousands of free
parameters \citep[e.g.][]{Galan2022herculens} are in that regime, and a Galaxy model with separate
disc, bulge and halo parameters fitted through a scalar likelihood would move the stellar-orbit
problem towards it; there the reverse-mode gradient of \gradsolve{} applies
\citep{SpurioMancini2026gradsolve}. Each thread keeps the state, stages and sensitivities of its trajectory in its own memory, which
suits the small systems of astrophysical ensembles; for larger states the library switches to a batched
\jax{} solver \citep{SpurioMancini2026gradsolve}. Problems with a large state, such as particle-mesh N-body
simulations \citep[e.g.][]{Modi2021flowpm, Li2024pmwd} or grid hydrodynamics
\citep[e.g.][]{Bezgin2023jaxfluids, Horowitz2025diffhydro}, are parallel within one trajectory, and
there a code that advances all components together as whole arrays is the natural design. In stiff
systems stability rather than accuracy sets the step of an explicit method
(Section~\ref{sec:adaptive}); in cosmology the Einstein--Boltzmann equations are an example, and
existing codes handle them with tight-coupling approximations or implicit solvers
\citep[e.g.][]{Lewis2000camb, Blas2011class, Sletmoen2026symboltz}. For stiff systems \gradsolve{}
provides a Rosenbrock kernel, benchmarked by \citet{SpurioMancini2026gradsolve} on two standard
stiff test problems, the Robertson chemical kinetics \citep{Robertson1966} and the HIRES
photomorphogenesis model \citep{Schafer1975hires}; \citet{HairerWanner1996} describe both.

All timings were made on one GPU model with right-hand sides in closed form. Because the
acceptance test~(\ref{eq:accept}) bounds the 99th percentile of the error, an analysis that needs
every trajectory within the accuracy requirement can apply the same test to the largest error instead;
the tables in the appendix list how many trajectories exceed the requirement for each code. The costs are given per evaluation or per iteration, the unit that enters optimisers, Fisher
forecasts and gradient-based samplers \citep[e.g.][]{Duane1987hmc, Neal2011hmc, Hoffman2014nuts} alike.

\section{Conclusions}
\label{sec:conclusions}

Many differentiable analyses in astrophysics solve a small system of ordinary differential equations
for thousands to millions of parameter sets. \gradsolve{} moves this computation to the GPU,
integrating each member of the ensemble in its own thread with its own step size and returning its
derivatives in the same pass. We measured the speed-up it provides in three examples, with every code held
to an accuracy requirement fixed in advance. For a million trajectories, \gradsolve{} is
\stNodeSpeedupGsEsix{} times faster than the serial CPU code running on all 128 cores of the
machine for the stellar orbits and their derivatives, where that code, \gala{}, is compiled C, and
\prNodeSpeedupGsEsix{} and \deNodeSpeedupGsEsix{} times faster for the spin precession and the
dark-energy distances, whose CPU codes, \precessionpkg{} and \scipy{}, evaluate a right-hand side written in Python; against a single
core the factors are larger again by \cpuNodeSpeedupLo{} to \cpuNodeSpeedupHi{} times. These ratios
combine hardware and software. On the
same GPU, \gradsolve{} is in addition about an order of magnitude faster than \diffrax{}, the
state-of-the-art ODE library in \jax{}, for large ensembles: the orbits and their derivatives are
obtained \stGradFactorDrawnEsix{} times faster, the dark-energy distances \deFwdFactorFullEsix{}
times faster including the solve for spatial flatness, and the precession \prFactorVernDrawn{}
times faster, for a million members each. For the stellar orbits this second step decides whether
the GPU pays off at all: with \diffrax{} the same GPU is slower than the 128 CPU cores. Where the
two codes integrate the same equations with the same Runge--Kutta method and tolerance, they take
almost the same number of steps, so the gap lies in the cost of each step. Inside the dark-energy
profile likelihood, a Gauss--Newton iteration becomes \deProfFactorFull{} times cheaper, and the
remaining cost lies in the likelihood itself rather than in the differential equation.

The approach suits many small, non-stiff systems differentiated with respect to a few parameters,
the setting of all three examples. For a scalar objective with many parameters the library's
reverse-mode gradient applies, systems with a large state suit codes that advance all components
together as whole arrays, and for stiff systems, such as those solved by Einstein--Boltzmann codes
\citep[e.g.][]{Lewis2000camb, Blas2011class}, the library provides a Rosenbrock kernel benchmarked on
standard stiff test problems \citep{SpurioMancini2026gradsolve}.

\section*{Acknowledgements}
ASM thanks Patrick Kidger for useful comments and acknowledges funding from a Leverhulme Trust
Research Leadership Award.

\section*{Data availability}
\gradsolve{} is publicly available at \url{https://github.com/ECLIPSE-AI4Science/gradsolve} and
can be installed with \code{pip install gradsolve}.

\bibliographystyle{mnras}
\bibliography{refs,refs_extra,refs_diff}

\appendix
\counterwithin*{equation}{section}
\renewcommand{\theequation}{\thesection\arabic{equation}}
\renewcommand{\theHequation}{app.\theequation}
\section{Numerical details}
\label{sec:app-numerics}

This appendix gives the equations behind Section~\ref{sec:method}: the Runge--Kutta step and its
error control, dense output, the derivatives, a count of step attempts for the two ways of
stepping through an ensemble, and the references, tolerances and timing procedure used in the
comparison.

\subsection{Runge--Kutta steps and error control}

An explicit Runge--Kutta step of length $h$ from $(t, y)$ forms $s$ intermediate states $Y_j$, each
from evaluations of $f$ at the states before it, and combines the same evaluations with two sets of
weights $b_j$ and $\hat b_j$:
\begin{equation}
\begin{gathered}
  Y_j = y + h\sum_{l<j} a_{jl}\, f(t + c_l h, Y_l, \theta), \qquad j = 1,\dots,s,\\
  y^{+} = y + h\sum_{j=1}^{s} b_j\, f(t + c_j h, Y_j, \theta), \\
  e = h\sum_{j=1}^{s} (b_j - \hat b_j)\, f(t + c_j h, Y_j, \theta).
\end{gathered}
\label{eq:rk}
\end{equation}
The coefficients $a_{jl}$, $b_j$, $\hat b_j$ and $c_j$ define the method. A method has order $\nu$
when the error it makes in one step shrinks as $h^{\nu+1}$, so that the error accumulated over a
fixed interval shrinks as $h^{\nu}$; Tsit5 and Dopri5 have $\nu=5$, Verner~7 has $\nu=7$ and Dopri8
has $\nu=8$. The weights $b_j$ and $\hat b_j$ give two estimates of the new state, of orders $\nu$
and $\nu-1$, from the same evaluations, so the error estimate $e \in \mathbb{R}^d$ costs no extra
evaluations of $f$. It is measured against the tolerances with the scaled norm
\begin{equation}
  \lVert e \rVert_{\rm tol} = \Bigl[ \frac{1}{d} \sum_{j=1}^{d} \Bigl(\frac{e_j}{s_j}\Bigr)^{2}
  \Bigr]^{1/2},
  \label{eq:errnorm}
\end{equation}
where $s_j = \code{atol} + \code{rtol}\,\max(|y_j|, |y_j^{+}|)$, and $y$ and $y^{+}$ are the
states at the start and end of the step. The step is accepted if
$\lVert e \rVert_{\rm tol} \le 1$ and is otherwise retried. In both cases the next step length is
\begin{equation}
  h_{\rm new} = h\,\min\bigl\{\alpha_{\max},\ \max\bigl[\alpha_{\min},\
  \varsigma\,\lVert e\rVert_{\rm tol}^{-1/\nu}\bigr]\bigr\},
  \label{eq:controller}
\end{equation}
which is the step that would just meet the tolerance, reduced by a safety factor $\varsigma<1$ and
kept between $\alpha_{\min}h$ and $\alpha_{\max}h$. \gradsolve{} uses $(\alpha_{\min},\alpha_{\max})
= (0.2, 5)$; the default \diffrax{} controller allows the step to grow by up to a factor of 10 and
does not shrink it after an accepted step.

\subsection{Dense output}

The evaluations of an accepted step also define a polynomial in $\sigma \in [0,1]$,
\begin{equation}
  y(t+\sigma h) \simeq y + h\sum_{j=1}^{s} b_j(\sigma)\, f(t + c_j h, Y_j, \theta),
  \label{eq:dense}
\end{equation}
with $b_j(1) = b_j$, which gives the solution anywhere inside the step. For Tsit5 an output at $t_k$ inside a step costs
one evaluation of this polynomial, with no extra evaluations of $f$. \gradsolve{} (Verner~7) instead repeats a shortened step from the start of the step to each output time.

\subsection{Derivatives}

When the $m$ differentiated parameters are shared by all members of the ensemble, as the halo
parameters of Section~\ref{sec:streams} are, the gradient of a scalar $\mathcal{L}$ built from the
outputs, such as a $\chi^2$, is
\begin{equation}
  \frac{\partial \mathcal{L}}{\partial\theta} = \sum_{i=1}^{n} \sum_{k=1}^{K}
  \frac{\partial \mathcal{L}}{\partial y_i(t_k)}\, S_i(t_k),
  \label{eq:chain}
\end{equation}
where $K$ is the number of output times and $S_i(t_k) \in \mathbb{R}^{d\times m}$. When every member
is a separate fit, as in the dark-energy minimisations of Section~\ref{sec:de-profile}, each member
has its own objective $\mathcal{L}_i$, whose gradient uses only its own $S_i(t_k)$ and has no sum over
$i$. The Jacobian of a vector of residuals is assembled from the same $S_i(t_k)$. In every case the
first factor comes from automatic differentiation of the code that builds $\mathcal{L}$ or the
residuals.

In equation~(\ref{eq:sens}) the matrix $J_y = \partial f/\partial y$ and $\partial f/\partial\theta$
are evaluated along the trajectory, and forward-mode automatic differentiation of $f$ supplies the
products $J_y S_i$ and $\partial f/\partial\theta$ at each intermediate state. The step length is
controlled with the norm~(\ref{eq:errnorm}) over the $d$ components of $y_i$ alone. Differentiating
the intermediate states of equation~(\ref{eq:rk}) with $h$ held fixed gives, with $S$ the
sensitivity at the start of the step,
\begin{equation}
  \frac{\partial Y_j}{\partial\theta} = S + h\sum_{l<j} a_{jl}\Bigl[J_y(Y_l)\,
  \frac{\partial Y_l}{\partial\theta} + \frac{\partial f}{\partial\theta}(Y_l)\Bigr],
  \label{eq:stage-sens}
\end{equation}
which is the same Runge--Kutta step applied to equation~(\ref{eq:sens}). On a given sequence of
steps the integrated $S_i$ is therefore exactly the derivative of the numerical solution, and the
polynomial of equation~(\ref{eq:dense}) applied to the sensitivities gives $S_i(t_k)$.

Reverse-mode differentiation of a scalar $\mathcal{L}(y(t_1))$ in its continuous form, the adjoint
method, integrates $\lambda(t) \in \mathbb{R}^d$ backwards along the forward solution:
\begin{equation}
\begin{gathered}
  \frac{\mathrm{d}\lambda}{\mathrm{d}t} = -J_y^{\mathsf T}\lambda, \qquad
  \lambda(t_1) = \Bigl(\frac{\partial \mathcal{L}}{\partial y(t_1)}\Bigr)^{\mathsf T}, \\
  \frac{\mathrm{d}\mathcal{L}}{\mathrm{d}\theta} = \lambda(t_0)^{\mathsf T}
  \frac{\partial y_0}{\partial\theta}
  + \int_{t_0}^{t_1}\lambda^{\mathsf T}\frac{\partial f}{\partial\theta}\,\mathrm{d}t.
\end{gathered}
\label{eq:adjoint}
\end{equation}
One backward integration thus gives the gradient of one scalar with respect to any number of
parameters \citep[e.g.][]{GriewankWalther2008}, and a Jacobian needs one backward integration per
residual. With $C_0$ the cost of the solve without derivatives and $R$ the number of scalars whose
gradients are needed, the two approaches cost about
\begin{equation}
  C_{\rm fwd} \approx (1+m)\,C_0, \qquad C_{\rm rev} \approx (1+\kappa R)\,C_0,
  \label{eq:disc-modes}
\end{equation}
where $\kappa$, the cost of one backward integration in units of $C_0$, is typically larger than
one, because the backward integration evaluates products of vectors with $J_y^{\mathsf T}$ and must
store or recompute the forward solution. A least-squares fit or a Fisher forecast needs the
derivatives of many residuals with respect to few parameters, which favours forward sensitivities; a
sampler that needs only the gradient of the log-likelihood has $R=1$ and favours reverse mode once
$m$ exceeds $\kappa$ \citep[see e.g.][]{Hindmarsh2005sundials, Ma2021sensitivity}.

\subsection{Counting step attempts}

Counting one step attempt of one trajectory as the unit of cost, and with $N_i$ the number of step
attempts of trajectory $i$, a batch that steps all trajectories together and a kernel with one
trajectory per thread cost
\begin{equation}
  C_{\rm batch} \propto n \max_{i} N_i, \qquad
  C_{\rm thread} \propto 32 \sum_{\mathcal{G}} \max_{i \in \mathcal{G}} N_i,
  \label{eq:cost}
\end{equation}
where $\mathcal{G}$ runs over warps of 32 consecutive trajectories. The ratio of the two is one when
all $N_i$ are equal and approaches $\max_i N_i/\bar{N}$, with $\bar{N}$ the mean, when each warp
holds trajectories with similar $N_i$; it therefore depends on the spread of the $N_i$ and on how
that spread falls across warps. Running the batch in chunks gives the second form with 32 replaced
by the chunk size. The count ignores the cost of one step attempt, the memory traffic and how fully
each code occupies the GPU, all of which enter the measured times.

\subsection{References, tolerances and timing}

The reference solution comes from \diffrax{} itself at a tight tolerance: \diffrax{} (Dopri8) at
\code{rtol} = \code{atol} = \stRefTol{} for the stellar orbits and \prRefTol{} for precession, and
\diffrax{} (Tsit5) at \code{rtol} = \deRefRtol{} and \code{atol} = \deRefAtol{} for dark energy. A
second solve with tolerances ten times larger (stellar orbits) or ten times smaller (the other two
examples) must agree with it to within a tenth of the accuracy requirement for every trajectory
checked (Appendix~\ref{sec:supp-reference}).

For the stellar orbits the error of the derivative is
\begin{equation}
  \epsilon_i^{S} = \frac{\lVert W_i \,(S_i - S_i^{\rm ref})\, D \rVert_{\rm F}}
  {\lVert W_i\, S_i^{\rm ref}\, D \rVert_{\rm F}},
  \label{eq:st-graderr}
\end{equation}
where $\lVert\cdot\rVert_{\rm F}$ is the Frobenius norm, $W_i$ is diagonal with entries $1/r_{E,i}$ for
the three position rows and $1/v_{E,i}$ for the three velocity rows, and $D = \mathrm{diag}(v_c, q)$,
which makes every entry dimensionless. One norm covers both columns because the $q$ column vanishes for an orbit in the plane, whereas the $v_c$ column never does.

Each code is run on a grid of tolerances spaced by a quarter of a decade, from loose to tight, and
the loosest tolerance that satisfies equation~(\ref{eq:accept}) on the timed ensemble is kept. For
the stellar orbits and precession \code{atol} equals \code{rtol}; for dark energy
$\code{atol} = \deAtolRatio{} \times \code{rtol}$. For \diffrax{} each explicit method is searched
in this way, at each chunk size, and the fastest accepted configuration is kept.

Each timing is one complete execution of the whole ensemble, with inputs and outputs on the GPU. Every
timed code is first called once untimed, which compiles it, and the timer then checks that no
compilation takes place inside the timed repetitions, so compilation is charged to neither code.
We quote the best of several repetitions and show their spread in Appendix~\ref{sec:supp-scaling}.
All GPU timings use the same GPU, one NVIDIA H200 NVL, and all arithmetic, on the GPU and the CPU, is in
double precision.

\section{Timing tables}
\label{sec:app-tables-sp}

Tables~\ref{tab:streams} and~\ref{tab:precession} give the costs and speed-up factors behind the
stellar-stream and spin-precession panels of Figure~\ref{fig:speedup}; Table~\ref{tab:de_forward}
in Appendix~\ref{sec:app-de} gives those for the dark-energy distances.

\begin{table}
\centering
\caption{Stellar-stream orbits on one H200 NVL: cost per orbit in $\mu$s for $n$ orbits. \emph{Forward}:
the final state of each orbit. \emph{Gradient}: the final state and its derivative with respect to
$(v_c, q)$. \emph{Fastest}: the fastest accepted \diffrax{} configuration (at
$n=10^5$ and $10^6$: forward \stFwdDfxArm{}, gradient \stGradDfxArm{}; at $n=10^4$: forward
\stFwdDfxArmEfour{}, gradient \stGradDfxArmEfour{}). \emph{Same}: \diffrax{} integrating the same sensitivity equations with the same
Runge--Kutta method and tolerance as \gradsolve{}. Both \diffrax{} rows take their fastest chunk
count. A factor is the
\diffrax{} cost over the \gradsolve{} cost (Section~\ref{sec:comparison}). Of the $10^6$ orbits,
\stFwdNover{} exceed the accuracy requirement for the final state with \gradsolve{} and
\stFwdDfxNover{} with the fastest accepted \diffrax{} configuration, and \stGradNover{} and
\stGradDfxNover{} for the derivative.}
\label{tab:streams}
\setlength{\tabcolsep}{3.5pt}
\begin{tabular}{lccc}
\toprule
 & $n=10^4$ & $10^5$ & $10^6$ \\
\midrule
\multicolumn{4}{l}{\emph{Forward}} \\
fastest \diffrax{} & \stFwdDfxDrawnEfour & \stFwdDfxDrawnEfive & \stFwdDfxDrawnEsix \\
\gradsolve{} & \stFwdGsDrawnEfour & \stFwdGsDrawnEfive & \stFwdGsDrawnEsix \\
factor & \stFwdFactorDrawnEfour & \stFwdFactorDrawnEfive & \stFwdFactorDrawnEsix \\
\midrule
\multicolumn{4}{l}{\emph{Gradient}} \\
fastest \diffrax{} & \stGradDfxDrawnEfour & \stGradDfxDrawnEfive & \stGradDfxDrawnEsix \\
\diffrax{}, same & \stGradSameDrawnEfour & \stGradSameDrawnEfive & \stGradSameDrawnEsix \\
\gradsolve{} & \stGradGsDrawnEfour & \stGradGsDrawnEfive & \stGradGsDrawnEsix \\
factor, fastest & \stGradFactorDrawnEfour & \stGradFactorDrawnEfive & \stGradFactorDrawnEsix \\
factor, same & \stGradSameFactorDrawnEfour & \stGradSameFactorDrawnEfive & \stGradSameFactorDrawnEsix \\
\bottomrule
\end{tabular}
\end{table}

\begin{table}
\centering
\caption{Spin precession for \prN{} binaries from $r=100\,M$ to $10\,M$ on one H200 NVL: cost per binary
in microseconds, and the factor of the fastest accepted \diffrax{} configuration over each \gradsolve{} method.
Of the \prN{} binaries, \prTsitNover{} exceed the accuracy requirement with \gradsolve{} (Tsit5) and
\prVernNover{} with \gradsolve{} (Verner~7). Both counts lie within what the acceptance test,
equation~(\ref{eq:accept}), allows.}
\label{tab:precession}
\begin{tabular}{lc}
\toprule
fastest \diffrax{} & \prDfxDrawn \\
\gradsolve{} (Tsit5) & \prTsitDrawn \\
\gradsolve{} (Verner~7) & \prVernDrawn \\
\midrule
factor, Tsit5 & \prFactorTsitDrawn \\
factor, Verner~7 & \prFactorVernDrawn \\
\bottomrule
\end{tabular}
\end{table}

\section{Dark-energy model and likelihood}
\label{sec:app-de}

\subsection{Equations of motion and flatness}

With the notation of Section~\ref{sec:de-model}, the full right-hand side of
equation~(\ref{eq:de-model}) is
\begin{equation}
\begin{aligned}
  \frac{\mathrm{d}u}{\mathrm{d}N} &= -\Bigl(3 + \frac{\mathrm{d}\ln E}{\mathrm{d}N}\Bigr)u
    - \frac{3\,v'(\phi)}{E^2}, \\
  \frac{\mathrm{d}\ln E}{\mathrm{d}N} &= -\frac{3\Omega_{\rm m}\mathrm{e}^{-3N}
    + 4\Omega_{\rm r}\mathrm{e}^{-4N}}{2E^2} - \frac{u^2}{2},
\end{aligned}
\label{eq:de-rhs}
\end{equation}
with $v'(\phi) = -\lambda v(\phi)$. The first line is the Klein--Gordon equation of the field in
e-fold time, the function $F$ of equation~(\ref{eq:de-model}). The Friedmann equation, the first
line of equation~(\ref{eq:de-model}), is solved for $E$ at every evaluation of the right-hand side,
so the right-hand side is a closed-form function of the state. The equation of state of the field
is
\begin{equation}
  w = \frac{E^2u^2/6 - v}{E^2u^2/6 + v},
  \label{eq:de-w}
\end{equation}
so a field at rest has $w=-1$. The start at redshift \deZstart{} is early enough that an earlier
start changes the distances by much less than the accuracy requirement.

The equation of state also explains why the amplitude has to be solved for. The Newton iteration
on equation~(\ref{eq:de-flat}) starts from $a_0 = 1 - \Omega_{\rm m} - \Omega_{\rm r}$, the
value for a field that never moves. Once the field moves, $w > -1$ and its energy density
$\rho_\phi = E^2u^2/6 + v(\phi)$ decreases,
\begin{equation}
  \frac{\mathrm{d}\rho_\phi}{\mathrm{d}N} = -3(1+w)\,\rho_\phi < 0,
  \label{eq:de-rho}
\end{equation}
so by today it is below its initial value $a_0$, and $E(0)^2 = \Omega_{\rm m} + \Omega_{\rm r} +
\rho_\phi(0) < 1$. The distances then differ from those of the flat model by a factor that changes
with redshift, so refitting $H_0$ does not absorb the difference. This is why $a_0$ has to be solved for, and why its derivative, equation~(\ref{eq:de-ift}), enters every iteration of the fit.

\subsection{Data model and minimiser}

The BAO data are the DESI DR2 measurements of $D_M/r_d$, $D_H/r_d$ and
$D_V/r_d = (z D_M^2 D_H)^{1/3}/r_d$ \citep{DESI2025dr2}, with the sound horizon $r_d$ from the
fitting formula of \citet{Aubourg2015} at a fixed baryon density. The supernova data are the
Pantheon+ magnitudes \citep{Scolnic2022pantheon, Brout2022}, whose distance modulus is computed from
the luminosity distance $D_L = (1+z)D_M$. In the profile likelihood the solver saves the state of
each cosmology at \deProfNodes{} output redshifts through its dense output, and the model distances
at the data redshifts are interpolated through these nodes by a cubic spline in $\ln(1+z)$.

The residuals are whitened: with $C^{1/2}$ the lower-triangular Cholesky factor of each data
covariance $C$, the residual vector is $r = C^{-1/2}(\text{model} - \text{data})$, and its
supernova part is projected orthogonally to $C^{-1/2}\mathbf{1}$, which is equivalent to
marginalising over the absolute magnitude of the supernovae with a flat prior. The quantity
minimised is then $\chi^2 = r^{\mathsf T}r$. At each fixed $\lambda$ it is minimised over
$\psi = (\Omega_{\rm m}, H_0)$ by the damped Gauss--Newton update
\begin{equation}
  \psi \leftarrow \psi - \bigl(J^{\mathsf T}J + \mu\,\mathrm{diag}\,J^{\mathsf T}J\bigr)^{-1}
    J^{\mathsf T}r, \qquad J = \frac{\partial r}{\partial\psi},
  \label{eq:de-gn}
\end{equation}
with a small fixed damping $\mu$ and the step shortened when it leaves a fixed trust region. The
$H_0$ column of $J$ is analytic, because distances scale as $1/H_0$ and $r_d$ depends on
$\Omega_{\rm m}h^2$. The $\Omega_{\rm m}$ column uses the derivative of the state along the flat
models, $\mathrm{d}y/\mathrm{d}\Omega_{\rm m}$, at every output redshift. With the solve for $a_0$
included, the Newton steps on equation~(\ref{eq:de-flat}) at each iteration start from the previous
amplitude moved to the new $\Omega_{\rm m}$ with $\mathrm{d}a_0/\mathrm{d}\Omega_{\rm m}$. Each
minimisation must pass the convergence test (\deProfConv{}), and $\chi^2$ is recomputed at the
parameters it returns.

\subsection{Timings}

Table~\ref{tab:de_forward} gives the cost of the distances for a population of cosmologies
(Section~\ref{sec:de-forward}). The fastest accepted \diffrax{} configuration is \deFwdDfxArm{}
with the amplitude supplied and \deFwdDfxArmFull{} including the solve for $a_0$; on the dark-energy
tolerance grid \code{atol} = \deAtolRatio{}\,\code{rtol}.

\begin{table}
\centering
\caption{Dark-energy distances: cost per cosmology in $\mu$s for $n$ cosmologies.
\emph{$a_0$ supplied}: the integration for a given potential amplitude. \emph{$a_0$ solved}: the full cost
from $(\lambda, \Omega_m)$, including the solve for $a_0$. \scipy{} (LSODA): one cosmology at a
time on one CPU core, timed with the amplitude supplied; its cost per cosmology is
independent of $n$, and its ratio to the GPU arms combines differences in hardware and software. \diffrax{}: the fastest accepted
configuration (\deFwdDfxArm{} with $a_0$ supplied, \deFwdDfxArmFull{} with $a_0$ solved). A factor is the \diffrax{} cost over the \gradsolve{} cost.
Of the $10^6$ cosmologies with the amplitude supplied, \deNover{} exceed the accuracy requirement
with \gradsolve{}. This count lies within what the acceptance test, equation~(\ref{eq:accept}),
allows.}
\label{tab:de_forward}
\setlength{\tabcolsep}{3.5pt}
\begin{tabular}{llccc}
\toprule
 & amplitude & $n=10^4$ & $10^5$ & $10^6$ \\
\midrule
\scipy{} (LSODA) & $a_0$ supplied & \multicolumn{3}{c}{\deFwdScipyCond} \\
fastest \diffrax{} & $a_0$ supplied & \deFwdDfxCondEfour & \deFwdDfxCondEfive & \deFwdDfxCondEsix \\
 & $a_0$ solved & \deFwdDfxFullEfour & \deFwdDfxFullEfive & \deFwdDfxFullEsix \\
\gradsolve{} & $a_0$ supplied & \deFwdGsCondEfour & \deFwdGsCondEfive & \deFwdGsCondEsix \\
 & $a_0$ solved & \deFwdGsFullEfour & \deFwdGsFullEfive & \deFwdGsFullEsix \\
\midrule
factor & $a_0$ supplied & \deFwdFactorCondEfour & \deFwdFactorCondEfive & \deFwdFactorCondEsix \\
 & $a_0$ solved & \deFwdFactorFullEfour & \deFwdFactorFullEfive & \deFwdFactorFullEsix \\
\bottomrule
\end{tabular}
\end{table}

Table~\ref{tab:de_profile} gives the cost of one Gauss--Newton iteration of the profile likelihood
(Section~\ref{sec:de-profile}) and how it divides between the differentiated solve and the rest of
the iteration, which includes whitening the residuals and solving equation~(\ref{eq:de-gn}) for the
update. The differentiated solve is \deProfSolveFactorCond{} times cheaper with \gradsolve{} than
with \diffrax{} when the amplitude is supplied and \deProfSolveFactorFull{} times cheaper with the
solve for $a_0$ included. On the mock data sets the fitted parameters of the two GPU arms agree to within
\deProfDthetaMax{} in $\Omega_{\rm m}$ or in $H_0$ (in km\,s$^{-1}$\,Mpc$^{-1}$), whichever differs
more, and at \deProfNcheck{} grid points the minima agree with those of the independent
minimiser (\scipy{}) to \deProfCheckMax{} in $\chi^2$. \gradsolve{} thus makes the differentiated
solve several times cheaper without changing the fits beyond these small differences.

\begin{table}
\centering
\caption{Profile likelihood of the dark-energy slope $\lambda$: cost of one Gauss--Newton iteration
per minimisation in $\mu$s, for $n=\deProfN{}$ minimisations on one H200 NVL. \emph{$a_0$ supplied} and \emph{$a_0$ solved} as in Table~\ref{tab:de_forward}. \scipy{} (LSODA): finite-difference Jacobians
on one CPU core. \diffrax{} (Tsit5): forward-mode differentiation of the batched solve. \gradsolve{}: sensitivities in the same launch as the solution. The middle rows split
the GPU iteration into the differentiated solve and the rest of the iteration, which both GPU
arms share. Factors are \diffrax{} over \gradsolve{}.}
\label{tab:de_profile}
\begin{tabular}{lcc}
\toprule
 & $a_0$ supplied & $a_0$ solved \\
\midrule
\scipy{} (LSODA) & \deProfScipyCond & \deProfScipyFull \\
\diffrax{} (Tsit5) & \deProfDfxCond & \deProfDfxFull \\
\gradsolve{} & \deProfGsCond & \deProfGsFull \\
\midrule
solve, \diffrax{} & \deProfDfxSolveCond & \deProfDfxSolveFull \\
solve, \gradsolve{} & \deProfGsSolveCond & \deProfGsSolveFull \\
rest of iteration & \deProfPostCond & \deProfPostFull \\
\midrule
factor, iteration & \deProfFactorCond & \deProfFactorFull \\
factor, solve & \deProfSolveFactorCond & \deProfSolveFactorFull \\
\bottomrule
\end{tabular}
\end{table}

\section{Supplementary material}
\label{sec:supplement}

\subsection{Converged references}
\label{sec:supp-reference}

The references are converged far beyond the accuracy requirements, so every error quoted for an arm
measures that arm rather than its reference. Every arm is judged against a reference computed with \diffrax{} (Dopri8) for the stellar orbits
and precession, or \diffrax{} (Tsit5) for dark energy, at the tight tolerances given in
Appendix~\ref{sec:app-numerics}. To show that the reference itself is converged, it is computed a
second time with \code{rtol} and \code{atol} a factor of ten apart, and the per-trajectory error
$\epsilon_i$ between the two, in the measure used for the arms, must stay below
$\epsilon_{\rm req}/10$, a tenth of the accuracy requirement. The check covers every orbit and every binary of the timed ensembles and the
first \deRefN{} cosmologies of each dark-energy ensemble. The largest difference, as a fraction of
$\epsilon_{\rm req}$, is \stRefDiffMax{} for the stellar orbits (final state and derivative),
\deRefDiffMax{} for the dark-energy distances and \prRefDiffMax{} for the precession angles. For the
profile likelihood on the observed data, the minima found by both GPU arms agree with the independent
minimiser (\scipy{}) to within \deObsIndepMax{} in $\chi^2$ at every value of $\lambda$ where
it was run.

\subsection{Work--precision and ensemble size}
\label{sec:supp-scaling}

\begin{figure*}
\begin{minipage}[t]{0.485\textwidth}
\centering
\includegraphics[width=\linewidth]{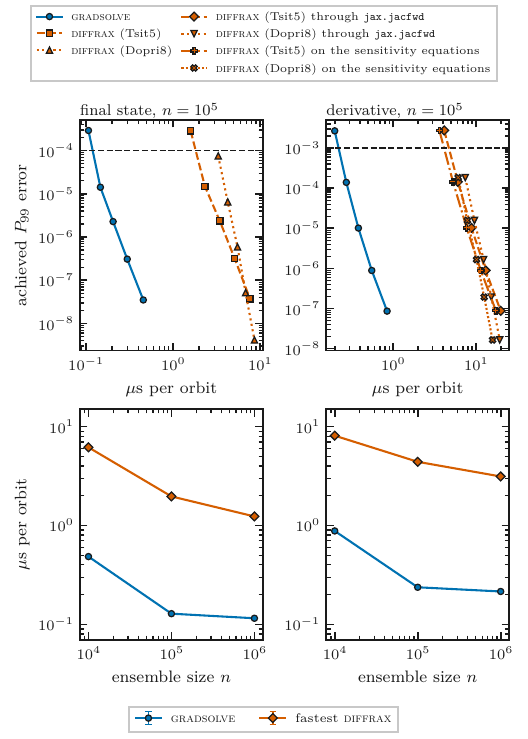}
\caption{Stellar-stream orbits, final state (left column) and its derivative with respect to
$(v_c, q)$ (right column). \emph{Top}: $P_{99}$ of the per-orbit error against cost per orbit at one
ensemble size, for \gradsolve{} and for each \diffrax{} candidate, \diffrax{} (Tsit5) and \diffrax{} (Dopri8), for
the derivative each through \code{jax.jacfwd} and on the sensitivity equations; one point per
tolerance; the dashed line marks the accuracy requirement. \emph{Bottom}: cost per orbit against $n$ for
\gradsolve{} and for the fastest accepted \diffrax{} configuration at each $n$; points are the best repetition and bars extend to the slowest.}
\label{fig:supp-streams}
\end{minipage}\hfill
\begin{minipage}[t]{0.485\textwidth}
\centering
\includegraphics[width=\linewidth]{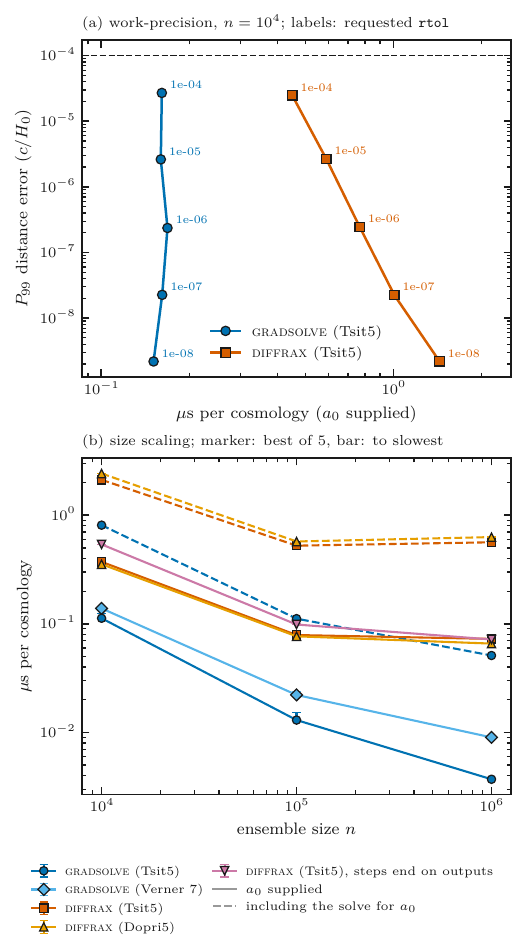}
\caption{Dark-energy distances at the \deNz{} DESI DR2 redshifts. (a) $P_{99}$ of the
per-cosmology distance error, in units of $c/H_0$, against cost per cosmology with the amplitude
supplied, for \gradsolve{} (Tsit5) and \diffrax{} (Tsit5), at one ensemble size; each point is
labelled with its requested \code{rtol}, and the dashed line marks the accuracy requirement. (b)
cost per cosmology against $n$ with the amplitude supplied (solid lines) for \gradsolve{} (Tsit5),
\gradsolve{} (Verner~7), \diffrax{} (Tsit5), \diffrax{} (Dopri5) and \diffrax{} (Tsit5) with its steps
ending on every output redshift instead of interpolating, and with the solve for $a_0$ included
(dashed lines) for \gradsolve{} (Tsit5), \diffrax{} (Tsit5) and \diffrax{} (Dopri5); markers are the best of five
repetitions and bars extend to the slowest.}
\label{fig:supp-darkenergy}
\end{minipage}
\end{figure*}

\begin{figure*}
\centering
\includegraphics[width=\textwidth]{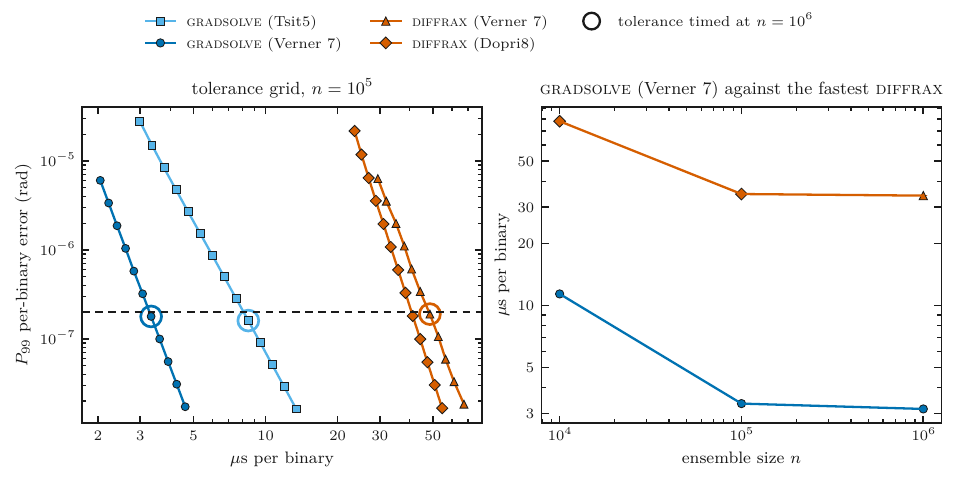}
\caption{Spin precession. \emph{Left}: $P_{99}$ of the per-binary angular error $\epsilon_i$
(equation~\ref{eq:prec-err}) against cost per binary on an independent draw, for
\gradsolve{} (Tsit5), \gradsolve{} (Verner~7) and the two \diffrax{} configurations that are fastest
accepted at some population size, \diffrax{} (Verner~7) and \diffrax{} (Dopri8), one point per tolerance; circles mark the tolerances timed on
the full population and the dashed line marks the accuracy requirement. \emph{Right}: cost per binary against
$n$ for \gradsolve{} (Verner~7) and for the fastest accepted \diffrax{} configuration
at each $n$ (marker shape as on the left); markers are the best repetition and bars extend to the
slowest.}
\label{fig:supp-precession}
\end{figure*}

A factor measured at one tolerance could favour one code if the two had different
work--precision relations, that is, different costs for the same achieved error. The upper or left
panel of Figures~\ref{fig:supp-streams}--\ref{fig:supp-precession} therefore plots the 99th percentile $P_{99}(\{\epsilon_i\})$ of the
per-trajectory error against the cost per trajectory as \code{rtol} and \code{atol} vary, with
the accuracy requirement marked. The lower or right panel plots the cost per trajectory against ensemble
size $n$ and shows the spread of the timing repetitions; the timing tables
(Tables~\ref{tab:streams}, \ref{tab:precession} and~\ref{tab:de_forward}) quote the best repetition.
In all three examples each \gradsolve{} curve lies at lower cost than each \diffrax{} curve over the
error range plotted, so the choice of tolerance does not reverse the comparison.

\end{document}